\documentclass[prb,twocolumn,%
               superscriptaddress, %
               groupedaddress, %
               amsfonts,amssymb,amsmath,%
               showpacs,%
               reprint ]{revtex4}
               
\usepackage[breaklinks=true]{hyperref}
\usepackage{latexsym}
\usepackage{verbatim} 
\usepackage{natbib}
\usepackage{graphicx}
\usepackage{textcomp}
\usepackage{xcolor}
\usepackage{xspace} 
\usepackage{ulem} 
\newcommand{\fhi}{Fritz-Haber-Institut der Max-Planck-Gesellschaft,14195, Berlin, Germany}
\newcommand{\dme}{Department of Mechanical Engineering and Material Science, Duke University, Durham, NC 27708, USA}
\newcommand{\tum}{Chair for Theoretical Chemistry, Technische Universit\"{a}t M\"{u}nchen, Lichtenbergstra{\ss}e 4, 85747 Garching, Germany}

\begin{document}
\title{The (3$\times$3)-SiC-($\bar{1}\bar{1}\bar{1}$) Reconstruction: Atomic Structure of the Graphene Precursor Surface from a Large-Scale First-Principles Structure Search}

\author{Jan Kloppenburg}
\thanks{Present address: Institute of Condensed Matter and Nanoscience, Universit\'e Catholique de Louvain,  Louvain-la-Neuve, 1348, Belgium}
\affiliation{\dme}
\affiliation{\fhi}

\author{Lydia Nemec}
\thanks{Present address: Carl Zeiss AG Digital Innovation Partners, Kistlerhofstraße 70, 81379 Munich, Germany}
\affiliation{\fhi}
\affiliation{\tum}

\author{Bj\"{o}rn Lange}
\thanks{Present address: DACS Laboratories GmbH, Niermannsweg 11-15, 40699 Erkrath, Germany }
\affiliation{\dme}

\author{Matthias Scheffler}
\affiliation{\fhi}

\author{Volker Blum}
\affiliation{\dme}

\date{\today}

\begin{abstract}
Silicon carbide (SiC) is an excellent substrate for growth and manipulation of large scale, high quality epitaxial graphene. On the carbon face (the ($\bar{1}\bar{1}\bar{1}$) or $(000\bar{1}$) face, depending on the polytype), the onset of graphene growth is intertwined with the formation of several competing surface phases, among them a (3$\times$3) precursor phase suspected to hinder the onset of controlled, near-equilibrium growth of graphene. Despite more than two decades of research, the precise atomic structure of this phase is still unclear. We present a new model of the (3$\times$3)-SiC-($\bar{1}\bar{1}\bar{1}$) reconstruction, derived from an {\it ab initio} random structure search based on density functional theory including van der Waals effects. The structure consists of a simple pattern of five Si adatoms in bridging and on-top positions on an underlying, C-terminated substrate layer, leaving one C atom per (3$\times$3) unit cell formally unsaturated. Simulated scanning tunneling microscopy (STM) images are in excellent agreement with previously reported experimental STM images.

\end{abstract}

\maketitle

\section{Introduction}

The ability to produce high-quality monolayer or few-layer graphene on a semiconducting substrate, as can be done on SiC,\cite{emtsev2009tws,Riedl_2009_H,riedl2010rev, deHeer2011pnas,emery2013cri,sforzini2015} remains a central question to fully leverage the properties of truly two-dimensional graphene. Indeed, commercially traded so-called graphene can be of questionable quality, composed of varying, relatively large numbers of graphene layers (often more than ten). \cite{boggild2018waronfakegraphene, kauling2018wwg} In contrast, graphene on Si-face SiC grows as a large-scale monolayer on its own by simply selecting the right thermodynamic conditions \cite{emtsev2009tws, riedl2010rev, deHeer2011pnas, emery2013cri, nemec2013tec, kunc2017eor}, and the growth of high-quality few-layer graphene is well documented on C-face SiC($\bar{1}\bar{1}\bar{1}$) \cite{deheer2011eg, berger2010epo, hass2008wmg}. Graphene grown on SiC thus avoid the problems of commercial graphene samples. 

On silicon carbide (SiC), high-quality graphene growth can be achieved by controlled sublimation of Si and thermal decomposition of the substrate.\cite{bommel1975laa, forbeaux2000ssg, berger2004ueg, kunc2017eor} The resulting, crystalline graphene layers are of exceptional quality, well suited for implementation of future graphene-based devices.\cite{berger2006eca, emtsev2009tws, lin2011wsg, her2012ttg, deheer2011eg, guo2013rmo}
The specifics of the graphene growth and properties depend strongly on the choice of the polar SiC surface face.\cite{riedl2010rev} For epitaxial graphene films on the Si-face of SiC ((111) for the 3C polytype, (0001) for hexagonal polytypes), large-area single- and few-layer graphene is achievable using growth conditions close to thermodynamic equilibrium.\cite{deHeer2011pnas,nemec2013tec} The atomic structure and homogeneity of these phases, including the subsurface region, is unambiguously confirmed, e.g., by x-ray standing waves,\cite{emery2013cri,sforzini2015} grazing-incidence x-ray diffraction\cite{schumann2013} or contact-resonance atomic force microscopy.\cite{Tu2016}
On the C-face (($\bar{1}\bar{1}\bar{1}$) or $(000\bar{1}$)), graphene layers show electrical properties similar to those of an isolated monolayer graphene film with very high electron mobilities.\cite{hass2008wmg, sprinkle2009fdo, berger2010epo} However, they are typically attributed to multilayer graphene growth. Monolayer graphene growth on the C-face has been attempted,\cite{mathieu2011mcb, hu2012seg, ruan2012ego} but not, to our knowledge, unambiguously achieved. A past paper by some of us\cite{nemec2015wgg} presented evidence that the growth mode on the C-face is affected by Si terminated surface reconstructions, which remain thermodynamically stable up to the C-rich chemical potential limit where bulk SiC itself is no longer stable against decomposition into bulk C (graphite). 

In a very recent publication \cite{Li_2019}, Li \textit{et al.} (including some of us) presented a more detailed study of surface phases occurring prior to the onset of near-equilibrium growth of graphene on the C-face of SiC without and with temperature effects and presence of hydrogen. In Ref.~\cite{Li_2019}, a preliminary description of a new (3$\times$3) surface reconstruction model was presented, which is known to occur along with graphene growth onset\cite{bommel1975laa, starke1997aso, hoster1997maa, li1996aso, bernhardt1999ssr, seubert2000iss, magaud2009got, hiebel2009aae, starke2009ego, hiebel2012sas} and the structure of which had previously eluded a definitive solution. 

Presenting the full computational evidence for and understanding of the structure of this key (3$\times$3) surface phase on C-face SiC is the main objective of this work. Specifically, we performed an {\it ab initio} random structure search (AIRSS) \cite{pickard2011air} to identify the nature of different (2$\times$2) and (3$\times$3) reconstructions of the SiC($\bar{1}\bar{1}\bar{1}$) surface at the onset of graphene growth. We note that further reconstructions with different periodicities were additionally addressed in Ref. \cite{Li_2019} 

\section{Background}

During annealing at high temperature (lower than but close to the Si sublimation limit, i.e., approaching the chemical potential limit favoring bulk graphite formation), a number of different surface structures are known to arise on the C-face of SiC.\cite{bommel1975laa, starke1997aso, hoster1997maa, li1996aso, bernhardt1999ssr, seubert2000iss, magaud2009got, hiebel2009aae, starke2009ego, hiebel2012sas} In the absence of oxygen and hydrogen, graphene growth starts with a Si rich (2$\times$2) phase with respect to the lateral unit cell of a hexagonal bilayer of bulk SiC.\cite{bernhardt1999ssr, li1996aso} Continued heating leads to a (3$\times$3) phase. Further annealing leads to a (2$\times$2) Si adatom phase, referred to as (2$\times$2)$_\text{C}$ (notation taken from \cite{bernhardt1999ssr}). Just before graphene forms on the surface, a coexistence of the (3$\times$3) and the (2$\times$2)$_\text{C}$ surface phases is observed. \cite{bernhardt1999ssr, veuillen2010iso}. While the atomic structure of the (2$\times$2)$_\text{C}$ reconstruction was resolved by quantitative low energy electron diffraction (LEED) \cite{seubert2000iss}, the nature of the (3$\times$3) reconstruction remained a puzzle. In the last decades, various structural models were suggested for the C-face (3$\times$3) phase.\cite{li1996aso, hoster1997maa, hiebel2012sas, deretzis2013adf, nemec2015wgg} Based on STM images, Li and Tsong proposed a tetrahedrally shaped cluster motif to explain the reconstruction.\cite{li1996aso} Hoster, Kulakov, and Bullemer suggested a different geometric configuration of atoms on the basis of STM measurements, albeit without specifying the chemical composition.\cite{hoster1997maa} Deretzis and La~Magna~\cite{deretzis2013adf} suggested a carbon rich composition with 6 C and 4 Si atoms. Hiebel \textit{et al.}~\cite{hiebel2012sas} suggested a model consisting of a SiC-bilayer with a stacking fault of one half of the cell and two adatoms, a Si adatom and C adatom. Experimentally, Auger electron spectroscopy (AES) indicates Si:C ratios of (6$\pm$2):1 \cite{hoster1997maa} and 1.2:1 \cite{bernhardt1999ssr}. Although they differ by a factor of $5\pm 2.67$, both experiments agree on a Si-rich near-surface composition. In the C-rich limit, this is initially somewhat surprising. However, in 2015, a computational analysis of several existing and new structure models by some of us \cite{nemec2015wgg} favored a model with all adatoms chosen to be Si, consistent with the high experimental Si:C ratios. The specific (3$\times$3) model discussed in Ref.~\cite{nemec2015wgg} was adapted from a well known (3$\times$3) reconstruction occurring on the Si-face.\cite{starke1998nrm} Qualitatively, the Si-rich termination may be understood by the greater stability of the heterogeneous Si-C bond over a homogeneous C-C bond -- thus, the dangling bonds of C atoms of the topmost Si-C bilayer show a preference for termination with Si, rather than C adatom motifs.
However, all (3$\times$3) models discussed in the past are still too high in energy, even at the chemical potential limit of bulk graphite formation, to coexist with the (2$\times$2)$_\text{C}$ adatom structure.\cite{nemec2015wgg} The structure model by Nemec \textit{et al.}~\cite{nemec2015wgg} reproduced most of the experimental characteristics listed above, but it failed to capture the characteristic STM images with the observed difference in intensity between occupied and empty state STM images (Fig.~2~II in Ref.~\cite{nemec2015wgg}).

\section{Method}

\begin{figure*}[ht]
  \centering
  \includegraphics[width=.98\textwidth]{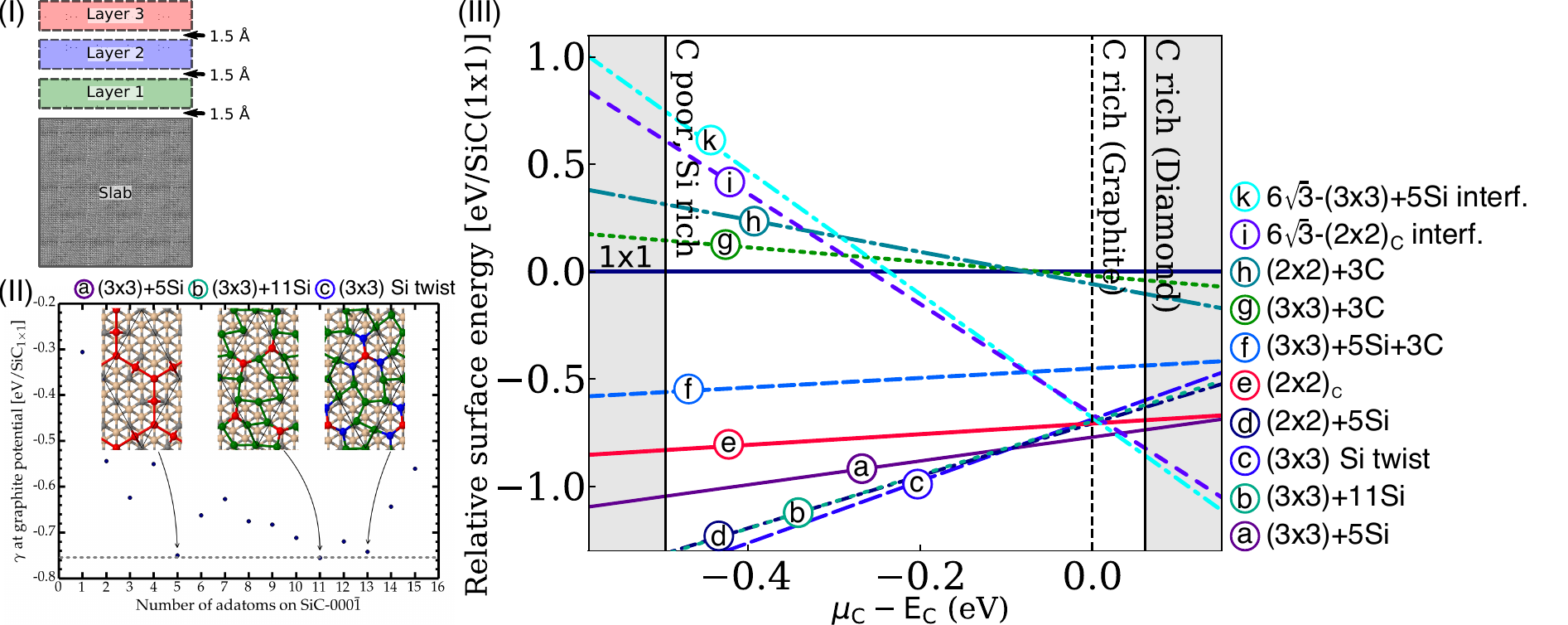}
  \caption{(I) Schematic view of the simulation slab. The random geometries are generated by distributing surface adatoms in three equidistant slabs. The slab coloring is the same for panel (I) and (II). (II) surface energies of the lowest energy structure found for 1 to 15 adatoms. (III) Surface energies as a function of the C chemical potential within the allowed ranges (given by diamond Si and graphite C). The surface energies are given relative to the bulk terminated (1$\times$1) surface. Shaded areas indicate chemical potential values outside the strict thermodynamic stability limits. (3$\times$3) AIRSS surface structures: (a) - (c) shown in panel (II), (c) AIRSS and Ref.\cite{nemec2015wgg}, (f) lowest energy structure for mixed Si and C adatoms, (g) lowest energy structure for C adatoms. (2$\times$2) AIRSS surface structures: (h) lowest energy structure for C adatoms, (e) AIRSS and Ref.\cite{seubert2000iss}, (d) lowest energy structure, (6 $\sqrt{3}\times$6$\sqrt{3}$) SiC/graphene interface structures: (i) $6\sqrt{3}$ interface structure with (2$\times$2)$_{\text{C}}$ structure (e) \cite{nemec2015wgg}, (k) $6\sqrt{3}$ interface with (3$\times$3) structure (a).}
  \label{fig1}
\end{figure*}

AIRSS has been successfully applied to a variety of systems \cite{pickard2010aat, pickard2011air, bogdanov2016mts}. Our AIRSS is based on van der Waals corrected density functional using the FHI-aims all-electron code \cite{blum2009aim,havu2009eoi,Levchenko_2015} employing the ELPA eigensolver library.\cite{auckenthaler2011pso, marek2014ELPA} Details of our AIRSS approach are listed Sec. II of the supplemental material (SM).\cite{supplementary} For the exchange-correlation functional, we use the van der Waals (vdW) corrected \cite{tkatchenko2009amv} Perdew-Burke-Enzerhof (PBE) generalized gradient approximation\cite{perdew1997gga} (PBE+TS) to calculate total energies. Unless otherwise noted, the calculations are non-spin-polarized. Technical parameters and bulk lattice constants are listed the SM, Secs. I.A. and I.B.\cite{supplementary} We restricted our search to the 3C-SiC polytype. In past work, details of reported surface reconstructions appeared to be unaffected by the detailed choice of polytype. \cite{furthmuller1998mot, schardt2000cot, pankratov2012goc, nemec2015wgg}

To identify the energetically most favorable reconstruction, we calculate the respective surface energies as formulated in the \textit{ab initio} atomistic thermodynamics approach.\cite{reuter2002csa} In the present paper, we neglect vibrational and configurational entropy contributions to the surface energy. The reconstruction models and energetics discussed here are therefore expected to hold for surface phases that form and stabilize at relatively low temperatures. Vibrational entropy effects that modify the higher-temperature equilibria are considered in more detail in Ref. \cite{Li_2019}. In the limit of sufficiently thick slabs, the surface energy $\gamma$ of a two-dimensional periodic SiC slab with a C face and a Si face is given as
\begin{eqnarray}\label{Eq:Esurf}
  \gamma & = & \gamma_\text{Si face} + \gamma_\text{C face} \\ 
  & = & \frac{1}{A_{\text{slab}}} \left(
  E^\text{slab} - N_\text{Si} \mu_\text{Si} -
  N_\text{C} \mu_\text{C} \right) \, . \nonumber
\end{eqnarray}
The letter $E$ denotes total energies for a given surface structure throughout this work. $N_\text{C}$ and $N_\text{Si}$ denote the number of C and Si atoms in the slab, respectively. Terminating H atoms at the bottom face (Si-face) of the slabs cancel in energy differences and are therefore not explicitly considered. We normalize the surface energy to the area of a (1$\times$1) SiC unit cell. The chemical potentials $\mu_\text{C}$ and $\mu_\text{Si}$ of C and Si are given by $\mu_\text{C} + \mu_\text{Si} = E^\text{bulk}_\text{SiC}$ where $E^\text{bulk}_\text{SiC}$ is the total energy of a bulk SiC unit cell (see Sec. I.C. of the SM\cite{supplementary}). The chemical potential limits of the C and Si reservoirs are fixed by the requirement that the underlying SiC bulk is stable against decomposition~\cite{nemec2013tec}, leading to:
\begin{equation}\label{Eq:muLimit}
   E^\text{bulk}_\text{SiC} - E^\text{bulk}_\text{Si} \le \mu_\text{C} \le E^\text{bulk}_\text{C}.
\end{equation}
We sampled the local energy minimum structures of the potential energy surface for the (2$\times$2) and (3$\times$3) surface slabs by calculating the surface energy in the carbon rich limit (Eqs.~\ref{Eq:Esurf}~and~\ref{Eq:muLimit}). Because of the close competition between the diamond and graphite structure for C \cite{berman1955,yin1984, choljun2014rpc}, we include both limiting phases in our analysis. The structure with the lowest energy for a given $\Delta \mu_\text{C}$  corresponds to the most stable phase.

The potential energy landscape is sampled by generating a large ensemble of random initial structures. These structures were created by distributing specific numbers of Si and C atoms randomly into up to three equidistant zones (``Layers'') above a C-terminated, 3-bilayer 3C-SiC($\bar{1}\bar{1}\bar{1}$) slab (see Fig.~\ref{fig1}(I) and the description in the SM, Sec. II.A.\cite{supplementary}). Each zone has a thickness of 2.5 {\AA}, separated by 1.5 {\AA} from adjacent zones. For the initial structures, minimum interatomic distances of 1.5 {\AA} were enforced. We limited the total number of randomly placed atoms in a single zone to nine for the (3$\times$3) superstructure searches and four for the (2$\times$2) searches. 
Overall, we calculated 38,447 structures with up to seven adatoms for the (2$\times$2) models and 21,008 structures with up to 15 adatoms for the (3$\times$3)  models (see SM Sec. II.~A.\cite{supplementary}). Specifically, we considered 2,097 C-only structures, 14,386 Si-only structures and 4,525 mixed Si+C structures for (3$\times$3) as well as 16,850 C-only structures, 7,197 Si-only structures and 14,430 mixed Si+C structures for (2$\times$2).
Each initial structure was then relaxed into their nearest local minimum at a ``light'' and computationally inexpensive level of theory, using three-bilayer SiC slabs (H-terminated at the bottom) as support and the local-density approximation (LDA) to density-functional theory for an initial rough energy weighting. Subsequently, a second relaxation at the PBE+TS level was performed at light settings. This led to groups of initial configuration converging to distinct local minimum energy structures (distinguished both by energy and atomic positions; see Fig.~2 in the SM \cite{supplementary}). Interestingly, in all cases considered, we found that the lowest-energy structure for a given composition was sufficiently energetically distinct from higher-energy structures (see Figs.~2 and 3 in the SM \cite{supplementary}). This allowed us to focus on only these lowest-energy structures for further processing atop six-bilayer SiC slabs using FHI-aims' ``tight'' settings at the PBE+TS level (for further details, see SM, Sec. I.A. \cite{supplementary}). 

For STM images, we used the SIESTA program\cite{SIESTA_2002} and utilities. Specifically, we employed the tool written by Pablo Ordej\'on and Nicolas Lorente\cite{siesta_stm_paper} and an additional tool to include the influence of the tip \cite{siesta_stm_workers}. We visualized the STM images using the WSxM\cite{wsxm_program} program. 

\section{Results}

\subsection{Lowest-energy surface reconstruction models}

In Figure~\ref{fig1}(II), we show the surface energies of the lowest energy structures found for one through 15 adatoms placed on the (3$\times$3) surface.
For the 13-adatom structure, our search confirmed the (3$\times$3) Si-twist model \cite{nemec2015wgg} as a low-energy structure (Fig.~\ref{fig1}(II), labeled \textit{c}). In addition, we identified two further low-energy structures, called (3$\times$3)+5Si (i.e.,  five Si adatoms, Fig.~\ref{fig1}(II), labeled \textit{a}) and (3$\times$3)+11Si (i.e., 11 Si adatoms, Fig.~\ref{fig1}(II), labeled \textit{b}). A top view of each of these three structures is depicted in the inset of Fig.~\ref{fig1}(II). Our search furthermore confirmed the (2$\times$2)$_\text{C}$ reconstruction \cite{seubert2000iss} for the (2$\times$2) surface (labeled \textit{e} in Fig.~\ref{fig1}(III)) and revealed an energetically lower, previously unknown 5-adatom structure for the (2$\times$2) surface (labeled \textit{d} in Fig.~\ref{fig1}(III)). This ``(2$\times$2)-5Si'' structure is shown in Fig. 6 in the SM and is discussed in greater detail in Ref.~\cite{Li_2019} as well. It is noteworthy that the (2$\times$2)$_\text{C}$ reconstruction proposed by Seubert \textit{et al.}~\cite{seubert2000iss}, labeled \textit{e}, would not become stable over the competing (3$\times$3) reconstructions (\textit{a} and \textit{c}) (see below) within the chemical potential limits. Interestingly, the lowest energy structures that we found contain only Si adatoms. Structures containing C adatoms are all to high in surface energy to be relevant.

The most promising search candidates for Si-only, C-only and mixed Si+C models of the (2$\times$2) and (3$\times$3) reconstruction are shown as a function of $\Delta \mu_\text{C} = \mu_\text{C}-E^\text{bulk}_\text{C}$ in Fig.~\ref{fig1}(III). In addition, two structures featuring graphene layers atop Si-rich reconstructions (labeled ``interf.'') that were constructed individually after the AIRSS was complete are also shown. We next discuss these alternative SiC-($\bar{1}\bar{1}\bar{1}$) surface phases (for more details see SM, Sec. III \cite{supplementary}). Two C-only adatom structures for the (3$\times$3) and (2$\times$2) reconstructions with the lowest surface energies are included in Fig.~\ref{fig1}(III), labeled \textit{g} and \textit{h}. For the (3$\times$3) periodicity, we also include a surface structure consisting of a mix of 5~Si and 3~C adatoms, labeled \textit{f}. The three structures labeled \textit{f, g} and \textit{h} are the structures with the lowest surface energies for models containing carbon atoms in the adlayer, unless considered as a full graphene layer. However, these structures are to high in surface energy to be relevant.

We next consider structures featuring only Si adatoms, which are the lowest-energy models for all chemical potentials in which a full graphene adlayer cannot yet exist. In the Si rich regime of the chemical potential, the most stable (3$\times$3) reconstruction is the Si-twist model \cite{nemec2015wgg}, labeled \textit{c}. For the carbon rich regime, we identify a new Si-rich adatom structure, namely the (3$\times$3)+5Si labeled \textit{a}. The Si-twist model has 13 Si adatoms and the (3$\times$3)+5Si has 5 adatoms, so the Si density of the Si-twist model surface is higher by a factor of 2.6 than in the (3$\times$3)+5Si model. Interestingly, the factor between these Si concentrations agrees well with the lower limit of factors found between the two past AES  measurements.\cite{hoster1997maa, bernhardt1999ssr} It is conceivable that the difference in the AES measurements could be explained by our surface diagram. While Hoster \textit{et al.}\cite{hoster1997maa} investigated the (3$\times$3) reconstruction in the Si-rich limit -- the Si-twist model, and Bernhardt \textit{et al.} \cite{bernhardt1999ssr} performed their measurements at the graphene precursor phase -- the (3$\times$3)+5Si. We also include a model for the (3$\times$3) reconstruction with 11 Si adatoms, labeled \textit{b}. Model \textit{b} consists of a Si ad-layer like the Si-twist model \textit{c}, but it differs in the second layer (see Fig~\ref{fig1}(II)). In summary, the main result of our search is the identification of the new (3$\times$3)+5Si adatom model that was not considered in any past work and that appears to fill a critical gap in understanding C-face surface phase formation on SiC immediately prior to the graphene growth onset.

\begin{figure}[ht]
  \centering
  \includegraphics [width=\columnwidth] {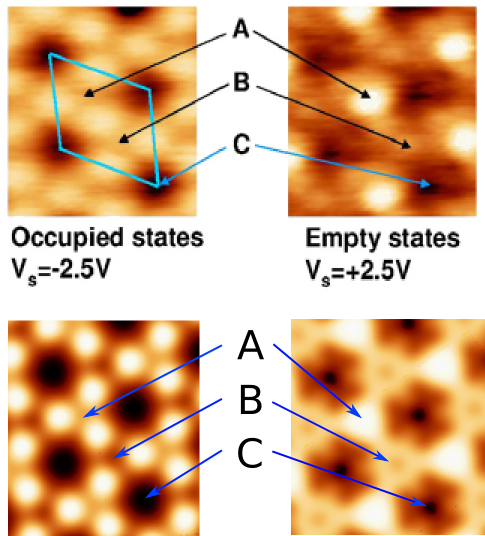}
\caption{Top: Experimental scanning tunneling microscopy (STM) images taken from Hiebel \textit{et al.}~\cite{hiebel2012sas} showing the characteristic change between occupied and unoccupied tip voltages. Bottom: Simulated constant-current STM images for occupied and empty states including a tip model.\cite{siesta_stm_workers} The scanning voltages are the same for the experimental and the simulated images. The three points of interest (A, B, C) marked by arrows are labeled according to Hiebel \textit{et al.}~\cite{hiebel2012sas}.
\label{fig2}}
\end{figure}

\subsection{Validation: Simulated STM and densities of states}

\begin{figure}[ht]
  \centering
  \includegraphics[width=\columnwidth]  {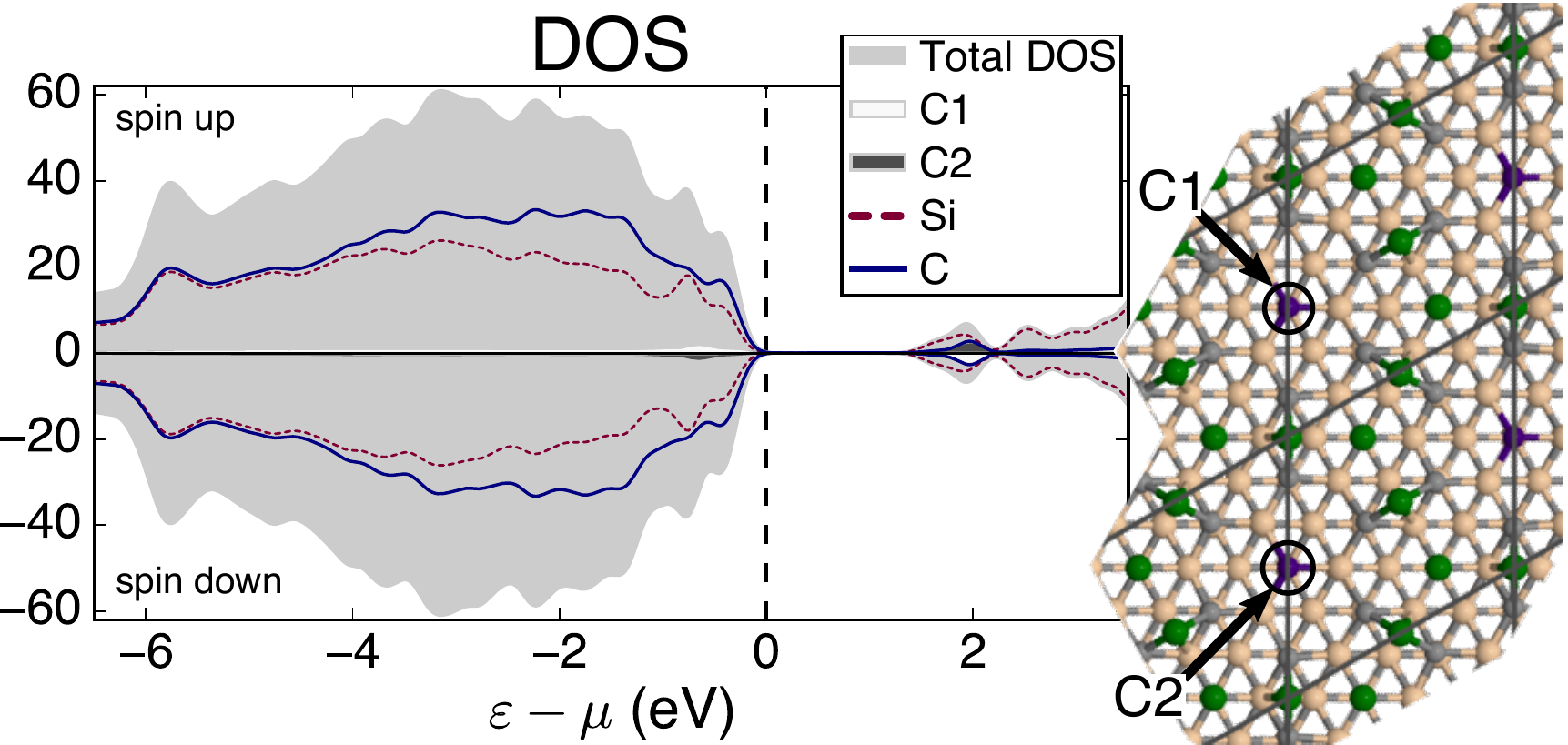}
\caption{Left: Total density of states (DOS) applying the HSE06 exchange-correlation functional and the projected DOS (p-DOS) projected on the Si, C and the two unsaturated carbon atoms (C1 and C2). Right: Image of the (2$\times$1) supercell of the (3$\times$3)+5Si surface reconstruction used for the spin-polarized simulation of the electronic structure. The unit cell is indicated in grey, the Si adatoms are marked in green and the carbon atoms of the p-DOS are highlighted in indigo.
\label{fig3}}
\end{figure}

A critical validation of the new (3$\times$3)+5Si adatom model can be obtained by comparing to existing scanning tunneling microscope (STM) images of the (3$\times$3) reconstruction,\cite{deretzis2013adf, hiebel2012sas}  which could not be explained by previous models. Specifically, the top row of Fig.\ref{fig2} reproduces  a characteristic difference between the empty and filled state images of the (3$\times$3) reconstruction as observed in past experimental work.\cite{deretzis2013adf, hiebel2012sas} 
Previous models failed to describe this difference in simulated STM images. In Figure~\ref{fig2}, we present a comparison of the experimental constant current STM images from Hiebel \textit{et al.}~\cite{hiebel2012sas} and simulated STM images of the (3$\times$3)+5Si model. In our simulation, we used the same voltage as in experiment for the occupied and empty states (-2.5~V and 2.5~V, respectively). In Figure~\ref{fig2}, the three characteristic points  (A, B, C) in the STM images are marked by arrows and labeled according to Hiebel \textit{et al.}~\cite{hiebel2012sas}. The simulated STM images of the (3$\times$3)+5Si reconstruction capture the specifics of the STM images very well. The triangular structure in the STM image (Fig.~\ref{fig2} bottom) had previously led to the assumption that the reconstruction should consist of a triangular adatom structure. To our surprise, structure \textit{a}, the (3$\times$3)+5Si surface model, does not feature such triangular arranged adatoms. It contains 5 Si adatoms organized in such a way that they form a net of equilateral hexagons on the surface with a side length of 5.34~\AA. In the middle of such a hexagon, we find there lies an unsaturated C atom (marked C1 and C2 in Fig.~\ref{fig3} right panel). We also show computed STM images for the alternative (3$\times$3) reconstruction models (3$\times$3)+11Si and (3$\times$3) Si-twist in Figures 13 and 14 in the SM.\cite{supplementary} The simulated images for these two reconstructions are markedly different from the experimental images and provide strong supporting evidence that the (3$\times$3)+5Si surface model is indeed observed in experiment.

Experimental scanning tunnelling spectroscopy (STS) of the (3$\times$3) reconstruction also shows a semiconducting surface with a $1.5$~eV band gap.\cite{hiebel2009aae} 
We calculated the spin-polarized electronic density of states (DOS) using the Heyd-Scuseria-Ernzerhof hybrid functional (HSE06) \cite{krukau2006iot} for the exchange-correlation functional. The DOS was calculated in a supercell of twice the (3$\times$3)+5Si simple unit cell, i.e., containing two unsaturated C atoms in the middle of the hexagon (see Fig.~\ref{fig3} right) and allowing the surface to assume a state with net zero spin polarization. In addition, we calculated the projected DOS (p-DOS) for all Si and C atoms in the simulation slab including the unsaturated C atoms labeled C1 and C2 in Fig.~\ref{fig3}.

Figure~\ref{fig3} shows the DOS and p-DOS for the spin down (negative $y$-axis) and spin up (positive $y$-axis) channel. The DOS clearly demonstrates that the surface is semiconducting, in agreement with experiment. It features a surface band gap of $1.67$~eV, which is smaller than the bulk 3C-SiC band gap of 2.51~eV on the same level of theory. However, it is in remarkable qualitative agreement with the band gap of $1.5$~eV from scanning tunnelling spectroscopy.\cite{hiebel2009aae} In the 3C-SiC bulk band gap close to the conduction band minimum, the surface reconstruction leads to one additional peak. The p-DOS indicates that this peak originates from the unsaturated C atoms (C1 and C2), while the corresponding occupied surface state is in resonance with the SiC bulk states. Thus, the (3$\times$3)+5Si appears to match the existing experimental observations very well.

\subsection{Relation to graphene-covered interface structures}

A final question concerns the interface structures covered by actual graphene layers on 
SiC($\bar{1}\bar{1}\bar{1}$), which have been documented by thorough STM studies. \cite{hiebel2012sas} From a theoretical viewpoint, graphene-covered (2$\times$2)$_{\text{C}}$ and (3$\times$3) Si-twist surface models in a (6$\sqrt{3} \times$6$\sqrt{3}$) supercell covered by a 13$\times$13 graphene monolayer were discussed in Ref.~\cite{nemec2015wgg}. Both models would not become stable in the thermodynamically allowed chemical potential range for C in our calculations, when compared to the new (3$\times$3)+5Si model. In Figure~\ref{fig1}(III), we include the surface energies of two graphene-covered surfaces with a nearly strain-free (6$\sqrt{3} \times$6$\sqrt{3}$)-R30$^\circ$ periodicity: the graphene-covered (2$\times$2)$_{\text{C}}$ from Ref.~\cite{nemec2015wgg}, labeled \textit{i}, and the graphene-covered (3$\times$3)+5Si interface structure, labeled \textit{k} (see SM, Sec. V, for details~\cite{supplementary}). The latter model, \textit{k}, is not thermodynamically stable in the allowed C chemical potential range, but it does become lower in energy than the non-graphene covered model very close past the limit of forming bulk graphite. Unlike the SiC(111) face, the structures considered in this work therefore do not indicate the existence of a thermodynamically stable interface between SiC($\bar{1}\bar{1}\bar{1}$) and a monolayer graphene phase. The existence of a different, stable graphene covered interface structure cannot be entirely ruled out by this limited set of structures. However, the crossover point to a graphene-covered phase observed by us would be consistent with the experimentally observed phase mixtures at SiC($\bar{1}\bar{1}\bar{1}$) surfaces close to graphitization, where graphene formation occurs kinetically only in the regime where bulk graphite is already the thermodynamically stable limit, consistent with the experimentally observed rapid formation of ``few-layer graphene''.

\section{Conclusion}

In this work, we address the long-standing open question of the SiC-$\bar{1}\bar{1}\bar{1}$-(3$\times$3) reconstruction. Our \textit{ab initio} random structure search reveals that the surface phase diagram of the SiC-$\bar{1}\bar{1}\bar{1}$ is even richer than previously anticipated. We identified two stable (3$\times$3) reconstructions, namely the previously published Si-twist model \cite{nemec2015wgg} in the Si rich regime of the chemical potential and the new (3$\times$3)+5Si in the C rich regime of the chemical potential. The difference in Si concentration is very similar to the difference between the published AES measurements. Additionally, we identified a new potential candidate for the experimentally observed (2$\times$2)$_{\text{Si}}$ reconstruction. We note that we restricted our structure searches to (3$\times$3) and (2$\times$2) periodicities, addressing the main cases reported in past experimental work but leaving room for other low-energy structures to exist in regimes not covered by these experiments. Indeed, two such phases of differing periodicity are identified in the recent computational study reported in Ref.~\cite{Li_2019}. Importantly, the (3$\times$3)+5Si model found and substantiated in the present work can reproduce experimental STM images and features a calculated (hybrid density functional theory) band gap of $1.67$~eV, in good agreement with the reported band gap in STS. Our structure search reveals that the C-face of SiC favors Si-only adatom structures over C-only or Si-C mixed adatom models. Overall, the study reported here as well as the recent study of Ref.~\cite{Li_2019}, which also considers vibrational entropy effects at finite temperature, provide a detailed picture of the critical atomic structures that govern the chemical potential onset of graphene formation on the C-face of SiC and a consistent framework for understanding key differences in graphene growth on Si-face versus C-face SiC.

\begin{acknowledgments}
L.N. acknowledge funding by the German Federal Ministry of Education and Research (BMBF) within project ``ELPA-AEO'' (project number 01IH15001). J.K. acknowledges a visiting student fellowship to Duke University, provided by the Fritz Haber Institute of the Max Planck Society (Berlin). We thank the DEISA Consortium, co-funded through the EU FP7 project RI-222919, for support within the DEISA Extreme Computing Initiative DECI-6. We furthermore thank Profs. Randall M. Feenstra, Michael Widom and their groups for stimulating discussions. 
\end{acknowledgments}

\bibliography{bibliography}
\end {document}


\normalem

\title{Supplemental Material: The (3$\times$3)-SiC-$\bar{1}\bar{1}\bar{1}$ Reconstruction: Atomic Structure of the Graphene Precursor Surface from a Large-Scale First-Principles Structure Search}

\author{Jan Kloppenburg}
\thanks{Present address: NAPS, Chemin des \'{E}toiles 8/L7.03.01, 
1348 Louvain-la-Neuve, Belgium }
\affiliation{\dme}
\affiliation{\fhi}

\author{Lydia Nemec}
\thanks{Present address: Carl Zeiss AG Digital Innovation Partners, Kistlerhofstraße 70, 81379 Munich, Germany}
\affiliation{\fhi}
\affiliation{\tum}

\author{Bj\"{o}rn Lange}
\thanks{Present address: DACS Laboratories GmbH, Niermannsweg 11-15, 40699 Erkrath, Germany }
\affiliation{\dme}

\author{Matthias Scheffler}
\affiliation{\fhi}

\author{Volker Blum}
\affiliation{\dme}

\date{\today}

\maketitle
%

We here discuss the technical details of our calculations of the surface phases of 3C-SiC($\bar{\text{1}}\bar{\text{1}}\bar{\text{1}}$) (C face). In particular, we discuss the details of the \textit{ab initio} random structure search. In addition, we discuss the computational details of the include simulated scanning tunneling microscope (STM) images.

\section{Computational details of the density-functional theory calculations}

\subsection{Electronic Structure: Total Energies, Forces, Densities of States}
In this work, we use the full potential, all electron code FHI-aims\cite{blum2009aim,havu2009eoi,Levchenko_2015} for density functional theory (DFT) based geometry optimization and ground state total energy and density of states (DOS) computations. (See Sec.~\ref{Sec:STM} for STM calculations, which were conducted with the SIESTA code.\cite{siesta_stm_paper,wsxm_program, SIESTA_2002, siesta_stm_workers}) FHI-aims employs numeric atom-centered basis sets.\cite{blum2009aim} We chose two levels of quality for the basis set and numerical real space grids. The \textit{light} quality settings include for Si a \textit{tier} 1-f and a \textit{tier} 1 basis set for C. The higher quality \textit{tight} settings include a \textit{tier} 1+dg for Si and a \textit{tier} 2 basis set for C.\cite{blum2009aim} Details about the numerical convergence with respect to the grid density in real- and reciprocal space and number of basis functions were also included in the supplemental material in a previous work\cite{nemec2013tec}.

For the exchange-correlation functional, we used both the local density approximation (LDA) functional parameterized by Perdew and Wang\cite{perdew1992aas} (for more rapid scanning of many structure models) and the generalized gradient approximation (GGA) parameterized by Perdew, Burke and Ernzerhof\cite{perdew1997gga} (PBE) with the van der Waals correction by Tkatchenko and Scheffler\cite{tkatchenko2009amv} (PBE+TS, for total-energy assessments of the identified structure models). We use the Heyd-Scuseria-Ernzerhof\cite{HSE2003} (HSE) hybrid functional in the flavour of 2006 (HSE06) with the suggested parameters\cite{krukau2006iot} $\alpha=0.25$ and $\omega=0.11~\text{Bohr}^{-1}$ and tight settings for DOS calculations. Unless otherwise noted the calculations are non-spin-polarized.

For our slab calculations, we employ $\Gamma$-centered k-point grids. Converged results were obtained by using a $k_x=k_y=6$, $k_z=1$ grid for the (2 $\times$ 2) models, and a $k_x=k_y=4$, $k_z=1$ for the (3 $\times$ 3) models. For the larger ($6 \sqrt{3} \times 6 \sqrt{3}$)-R30$^\circ$ interfaces, $\Gamma$-point calculations suffice.

All final surface structures are calculated using a slab of six SiC bilayers. The bottom silicon atoms are hydrogen terminated. The top three SiC bilayers and all adatoms or planes above are fully relaxed (residual energy gradients: $8 \cdot 10^{-3}$~eV/{\AA} or below).

\subsection{Bulk Structure and Enthalpy of Formation of \textit{3C}-SiC, Diamond, Graphite and Silicon}\label{ssec:bulkstructer}

In Table~\ref{tab:lattice_para_bulk}, we list the calculated optimized lattice parameters $a_0$ and cohesive energies of the reference structures 3C-SiC, silicon, and for the carbon references diamond and graphite, as well as the SiC enthalpy of formation, all at the PBE+TS level of theory. We do not consider zero-point or other vibrational corrections to the free energies of the solid phases since the dominant free-energy effect is expected to enter through the reservoir chemical potentials (see, however, Ref.~\cite{Li_2019} for a detailed computational assessment of vibrational free energy effects). The enthalpy of formation of a compound from different bulk elemental species $i_\text{species}$ is defined as 
\begin{equation}
    \Delta H_f = E_{\text{tot}} - \sum\limits_{i_\text{species}}{E_{\text{bulk}}(i_\text{species})}
\end{equation}
and the cohesive energies are defined as
\begin{equation}
   E_{\text{coh}} = \frac{1}{N_\text{tot}} \left(E_\text{tot} - \sum\limits_{i_\text{species}}{E_{\text{atom}}(i_\text{species})}\right) .
\end{equation}

\begin{table}[h]
 \centering

\begin{tabular}{l l| l}
\hline
  && PBE+TS\\
  \hline
3C-SiC &$a_0$ [\AA] & 4.36 \\
&$E_{\text{coh}}$ [eV/atom]&-6.76  \\
&$\Delta H_{f}$ [eV] & -0.56  \\
\hline
Silicon &$a_0$ [\AA] & 5.45 \\
&$E_{\text{coh}}$ [eV/atom] & -4.87  \\
\hline
Diamond &$a_0$ [\AA] & 3.55 \\
&$E_{\text{coh}}$ [eV/atom] &-7.93 \\
\hline
Graphite & $a_0$, $c_0$ [\AA] & 2.46, 6.64\\
&$E_{\text{coh}}$ [eV/atom]  &-8.00 \\
\hline
 \end{tabular}
 \caption{\label{tab:lattice_para_bulk} Lattice parameter $a_0$ and $c_0$ and cohesive energies $E_{\text{coh}}$ for 3C-SiC and silicon in diamond structure and for the carbon references diamond and graphite. The experimental lattice parameter of 3C-SiC is 4.36 {\AA}.\cite{li1986teo} $\Delta H_{f}$ denotes the enthalpy of formation  of 3C-SiC.}
\end{table}

\subsection{Surface Energies and Chemical Potentials}

The possible equilibrium conditions for different surface phases can be represented by the chemical potentials of C and Si, $\mu_\text{C}$ and $\mu_\text{Si}$ in a grand canonical formalism.\cite{reuter2002csa} In the limit of sufficiently thick SiC slabs the surface energies $\gamma$ of a two-dimensional periodic surface with a C face and a Si face is defined as
\begin{equation}\label{eq:surf_energy}
  \gamma=\gamma_\text{Si-face} + \gamma_\text{C-face} = \frac{1}{A_{slab}} \left(
  E^\text{slab} - N_\text{Si} \mu_\text{Si} -
  N_\text{C} \mu_\text{C} \right) \, ,
\end{equation}
where $N_a$ denotes the total number of atoms and $\mu_a$ denotes the chemical potential of atom type $a$. The term $A_{slab}$ represents the surface area of the slab supercell. $E_{\textrm{slab}}$ is the total energy of the slab. In our calculations, we choose a fixed H-terminated Si-face geometry at the bottom of the slab, which cancels out for all surface energy differences related to the C face.

The major experimental temperature and pressure ($T,p$) dependence during growth arises through the reservoirs of Si and C, which define $\mu_\text{Si}$ and $\mu_\text{C}$.\cite{reuter2002csa,tromp2009tak} The chemical potentials are limited by carbon rich $(\mu_C\geq E^{Bulk}_\textrm{Graphite}$) or carbon poor $(\mu_{Si}\geq E^{Bulk}_\textrm{Silicon})$ growth conditions. For the carbon rich limit we include the diamond and graphite structure in our surface energy phase diagrams. Using PBE+TS total energies, graphite is more stable than the diamond phase by 67~meV per atom. For the structure search, we use the energy of carbon in the graphite structure. 
The chemical potentials are thus obtained from
\begin{align}
a)\,\mu_\textrm{C} = \frac{E^{Bulk}_\textrm{Graphite}}{4}\hspace*{4mm}
b)\,\mu_\textrm{Si} = \frac{E^{Bulk}_\textrm{Silicon}}{2}\hspace*{4mm}
\end{align}

In thermodynamic equilibrium the chemical potentials of silicon
and carbon in are related by
\begin{equation}\label{eq:sic_mu}
E^{\textrm{SiC}}=2\left(\mu_{\textrm{C}}+\mu_{\textrm{Si}}\right)
\,\,\rightarrow\,\,
\mu_{\textrm{Si}}=\frac{E^{\textrm{SiC}}}{2}-\mu_{\textrm{C}}\, .
\end{equation}

\section{\textit{Ab initio} Random Structure Search (AIRSS)}
{\textit Ab initio} random structure searches (AIRSS) have been described extensively in literature.\cite{catlow2010, pickard2010aat, pickard2011air, woodley2014} In this section, we will describe the details of our present AIRSS.

\subsection{Random Geometry Generation and Geometry Optimization}

The potential energy landscape is sampled by generating a large ensemble of random initial structures. The slab geometry is generated from the 3C-SiC bulk structure from Sec.~\ref{ssec:bulkstructer}. 
The constructed (1$\times$1) slab geometry was relaxed with tight settings and PBE+TS. From this (1$\times$1) slab, we build (2$\times$2) and (3$\times$3) slabs by repetition of the in-plane lattice vectors.

\begin{figure}[ht]
 \includegraphics[width=0.25\textwidth]{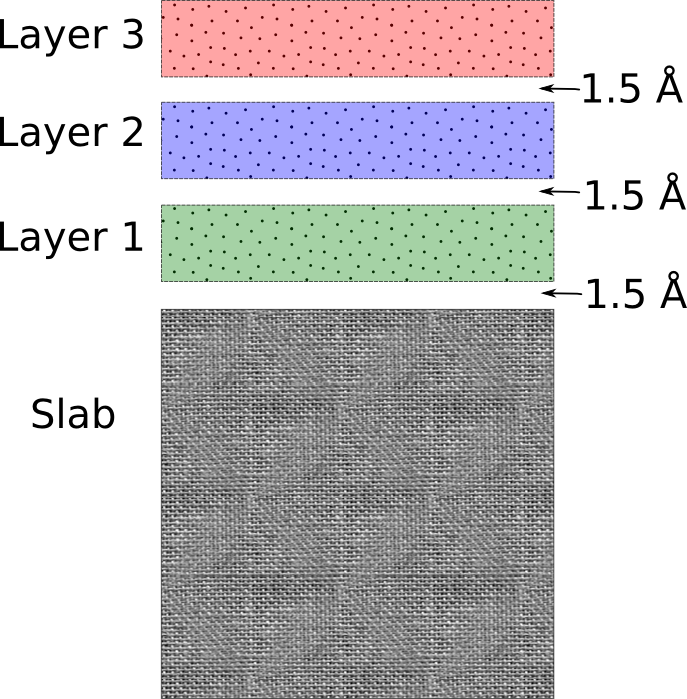}
 \caption{Schematic view of generating random geometries on a predefined slab -- see text for description. The layers are colored the same as later in our geometry pictures showing the found low energy candidates.\label{fig:substrate_with_layers}}
\end{figure}

The random structures were generated by distributing the desired number of Si and C atoms in up to three equidistant zones above a C-terminated three-bilayer slab (schematically illustrated in Fig.~\ref{fig:substrate_with_layers}). Each zone has a thickness of 2.5 {\AA} and a distance of 1.5 {\AA} to the next one below. To avoid generation of unphysical models, the $z$ component of adatoms is generated within a chosen layer thickness, measured from the top of the SiC($\bar{1}\bar{1}\bar{1}$). Minimum interatomic distances of 1.5~{\AA} were enforced.

We chose a two-step optimization procedure. First, the random initial geometry atop a three-SiC-bilayer slab was relaxed using light settings and the LDA functional (used here to achieve greater computational speed). Second, we used the converged structure and performed an additional relaxation using tight computational settings and the PBE+TS functional. Ultimately, we transfer the relaxed adatom structure onto a six-bilayer SiC slab and perform a final relaxation to obtain the total energy using tight computational settings and the PBE+TS functional. To further minimize the computational effort and to simulate bulk-like conditions, we keep the hydrogen-terminated bottom bilayer as well as one (two) additional bilayers above it fixed for the three (six) bilayer slabs, respectively.

The maximum number of randomly distributed adatoms per layer (cf. Fig.~\ref{fig:substrate_with_layers}) above the (2$\times$2) surface is four and above the (3$\times$3) surface it is nine. These values correspond roughly to one monolayer of atoms atop the respective surfaces. The maximum number of adatoms and the total number of structures considered in our AIRSS are listed in Table~\ref{tab:number_slabs}.

\begin{table}[ht]
\begin{tabular}{l|ll|ll}
\hline
\multirow{2}{*}{Surface} & \multicolumn{2}{c|}{(2$\times$2)} & \multicolumn{2}{c}{(3$\times$3)} \\
   & \# adatoms & $\sum$ structures & \# adatoms & $\sum$ structures \\
\hline
C-only      & 7 & 16850 & 15 & 2097  \\
Si-only     & 7 & 7197  & 15 & 14386 \\
C+Si        & 5 & 14430 & 9  & 4525  \\
\hline
Total       &   & 38477 &    & 21008  \\  \hline
\end{tabular}
\caption{The maximum number (\#) of adatoms and ($\sum$) total amount of randomly generated structures for the (2$\times$2) and (3$\times$3) SiC-$\bar{1}\bar{1}\bar{1}$ for C and Si adatoms only and C + Si chemical compositions.\label{tab:number_slabs}}
\end{table}

\subsection{Exploring the Energy Landscape within a Random Search}

\begin{figure}[h]
 \includegraphics[width=0.45\textwidth]{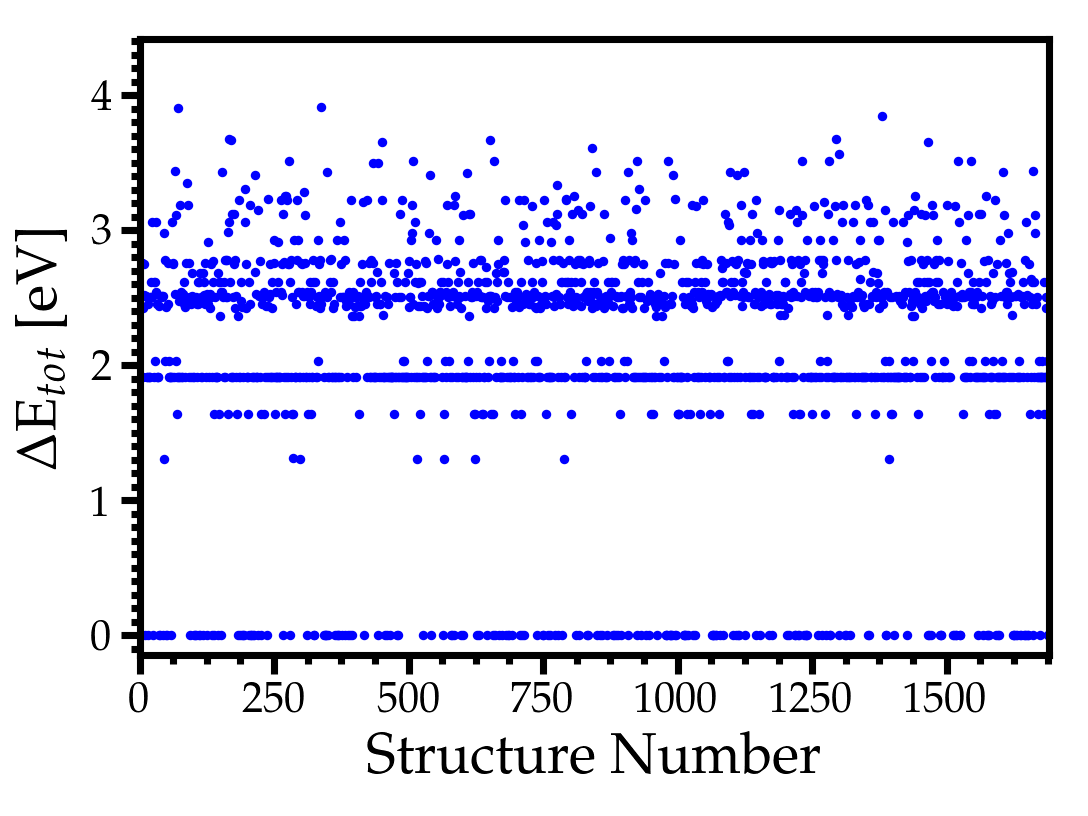}
 \caption{Energies of the structures resulting from the (3$\times$3)~+~5~Si random structure search, covering 1,689 random structures. An energetic ordering clearly emerges in the visual form of horizontal lines around prominent local energy minima. The overall lowest total energy was taken as the reference value and subtracted from each individual total energy.
 \label{fig:3x3+5_dotplot}}
\end{figure}

To obtain an impression of the energy landscape of different locally stable structures emerging from the AIRSS, we plotted the final total energy of each single geometry relaxation against its ``structure number'' in the search.  As an example, we show this ``dot-plot'' of the random search for the adatom configuration (3$\times$3)~+~5~Si (see Fig.~\ref{fig:3x3+5_dotplot}). The results reflect three-bilayer SiC slabs and light computational settings.

\begin{figure}[h]
 \includegraphics[width=0.45\textwidth]{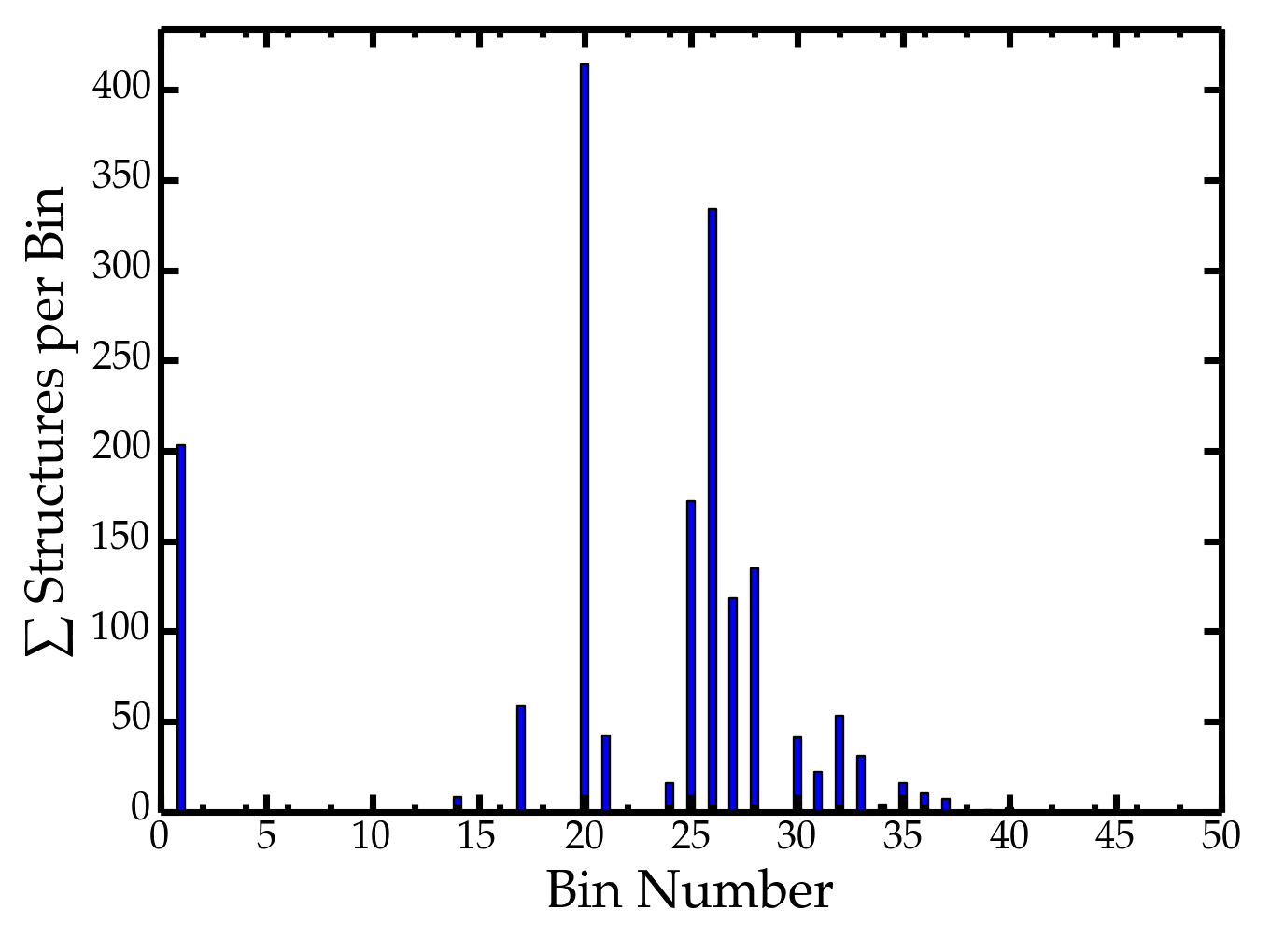}
 \caption{Energy histogram of the (3$\times$3)~+~5~Si random search results. The entire energy range of 3.92 eV is divided into 50 bins of 78~meV bin width each.
 \label{fig:3x3_binplot}}
\end{figure}

The energy landscape presented by dot-plots gives an impression of whether the number of generated structures was sufficient for finding a lowest energy candidate and how often structures occur that are energetically close to each other. The dot-plot of the random structure search for the (3$\times$3)~+~5~Si covered a total of 1,689 random structures. Figure~\ref{fig:3x3+5_dotplot} clearly reveals horizontal lines that indicate local minimum-energy structures in the energy landscape. In all dot-plots shown in this work, the overall lowest total energy has been taken as a reference value and subtracted from each individual result.

For the (3$\times$3)~+~5~Si random search results, we additionally show a histogram plot of the frequency of finding structures with similar energies in Fig.~\ref{fig:3x3_binplot}. The full energy range from 0 to 3.92~eV was divided into 50 bins with a bin width of 78~meV. For the first bin, the structures lay within an energetic window of 10~meV total energy deviation.

Similar protocols to the above description were employed for all other AIRSS searches in this work. As a second important point, we note that the random structure search of the (3$\times$3)~+~13~Si confirmed the previously published (3$\times$3)~Si-twist structure by Nemec \textit{et al.}~\cite{nemec2015wgg}. We included a total of 467 random structures for the (3$\times$3)~+~13~Si structure.

\section{The SiC($\bar{1}\bar{1}\bar{1}$) Surface Reconstructions}

In our \textit{ab initio} random structure search, we included two surface periodicities of the 3C-SiC($\bar{\text{1}}\bar{\text{1}}\bar{\text{1}}$) (C face): (3$\times$3) and (2$\times$2). The chemical composition of the surface reconstructions is not \textit{a priori} clear for either case. While the (2$\times$2)$_\text{C}$ reconstruction was resolved by quantitative low energy electron diffraction (LEED) \cite{seubert2000iss}, the (3$\times$3) reconstruction remains a puzzle. However, in addition to the (2$\times$2)$_\textrm{C}$ surface reconstruction a Si rich (2$\times$2) phase has been reported to appear before graphene growth starts.\cite{bernhardt1999ssr, li1996aso} For both structure searches, we therefore scanned different chemical compositions, searching for Si- and C-only reconstructions as well as Si+C mixed ones. 
As a reminder, the clean SiC($\bar{1}\bar{1}\bar{1}$) surface is terminated by four carbon atoms for the (2$\times$2) and nine carbon atoms for the (3$\times$3) periodicity, each with one unsaturated bond. 

\subsection{{The 3C-SiC($\bar{\text{1}}\bar{\text{1}}\bar{\text{1}}$)-(2$\times$2) Surface Reconstruction}}

Tables~\ref{tab:2x2_one_geo_run} and \ref{tab:2x2_mix_geo_run} show a summary of the total number of geometries that were generated and relaxed and the number and type of adatoms that were added to the surface for our AIRSS.

\subsubsection{3C-SiC($\bar{\text{1}}\bar{\text{1}}\bar{\text{1}}$)-(2$\times$2) Si-only and C-only}

\begin{table}[ht]
 \begin{tabular}{c|c|c||c|c|c}
  \hline
  \hline
  Element & adatoms & \# & Element & adatoms & \#\\
  \hline
   C & 1 & 3,767 & Si & 1 & 980\\
   C & 2 & 2,936 & Si & 2 & 904\\
   C & 3 & 2,543 & Si & 3 & 1,467\\
   C & 4 & 2,474 & Si & 4 & 1,117\\
   C & 5 & 2,007 & Si & 5 & 939\\
   C & 6 & 1,608 & Si & 6 & 950\\
   C & 7 & 1,515 & Si & 7 & 840\\
  \hline
  \hline
 \end{tabular}
 \caption{Total number of relaxed structures for the (2$\times$2) 3C-SiC($\bar{\text{1}}\bar{\text{1}}\bar{\text{1}}$) surface reconstruction with given numbers of Si-only or C-only adatoms.\label{tab:2x2_one_geo_run}}
\end{table}

\begin{figure}[ht]
 \includegraphics[width=0.45\textwidth]{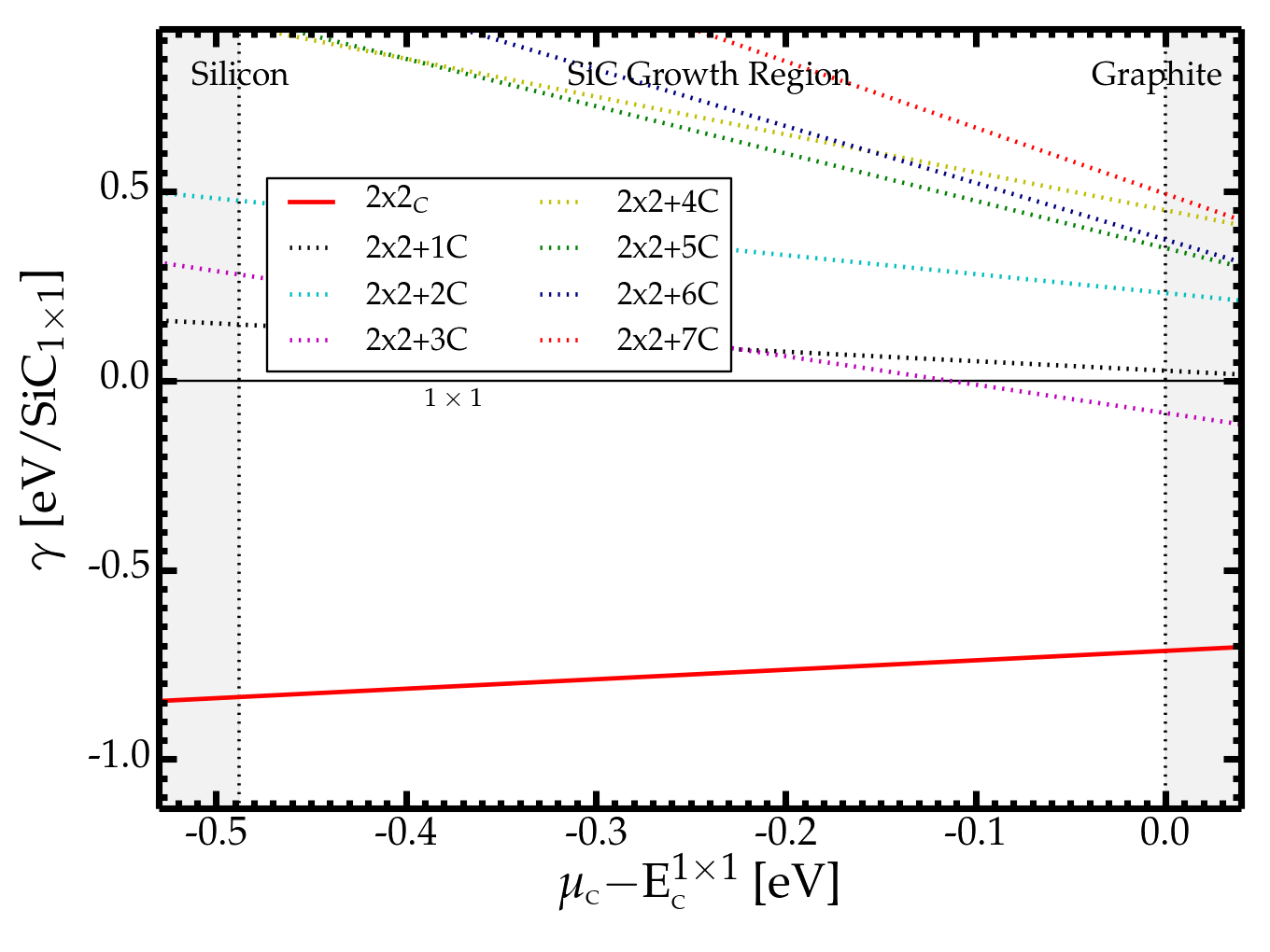}
 \caption{Results for the (2$\times$2) 3C-SiC($\bar{\text{1}}\bar{\text{1}}\bar{\text{1}}$) surface reconstruction composed of carbon adatoms only.
 \label{fig:2x2_carbon}}
\end{figure}

There were no energetically favorable results from our random searches when adding only carbon onto the slab. The results are shown in the surface phase diagram for adding carbon adatoms only in Fig.~\ref{fig:2x2_carbon}. The surface energies shown in Fig.~\ref{fig:2x2_carbon} were calculated using Eq.~\ref{eq:surf_energy}, tight settings with PBE+TS and a six-bilayer SiC slab.

\begin{figure}[ht]
  \includegraphics[width=0.45\textwidth]{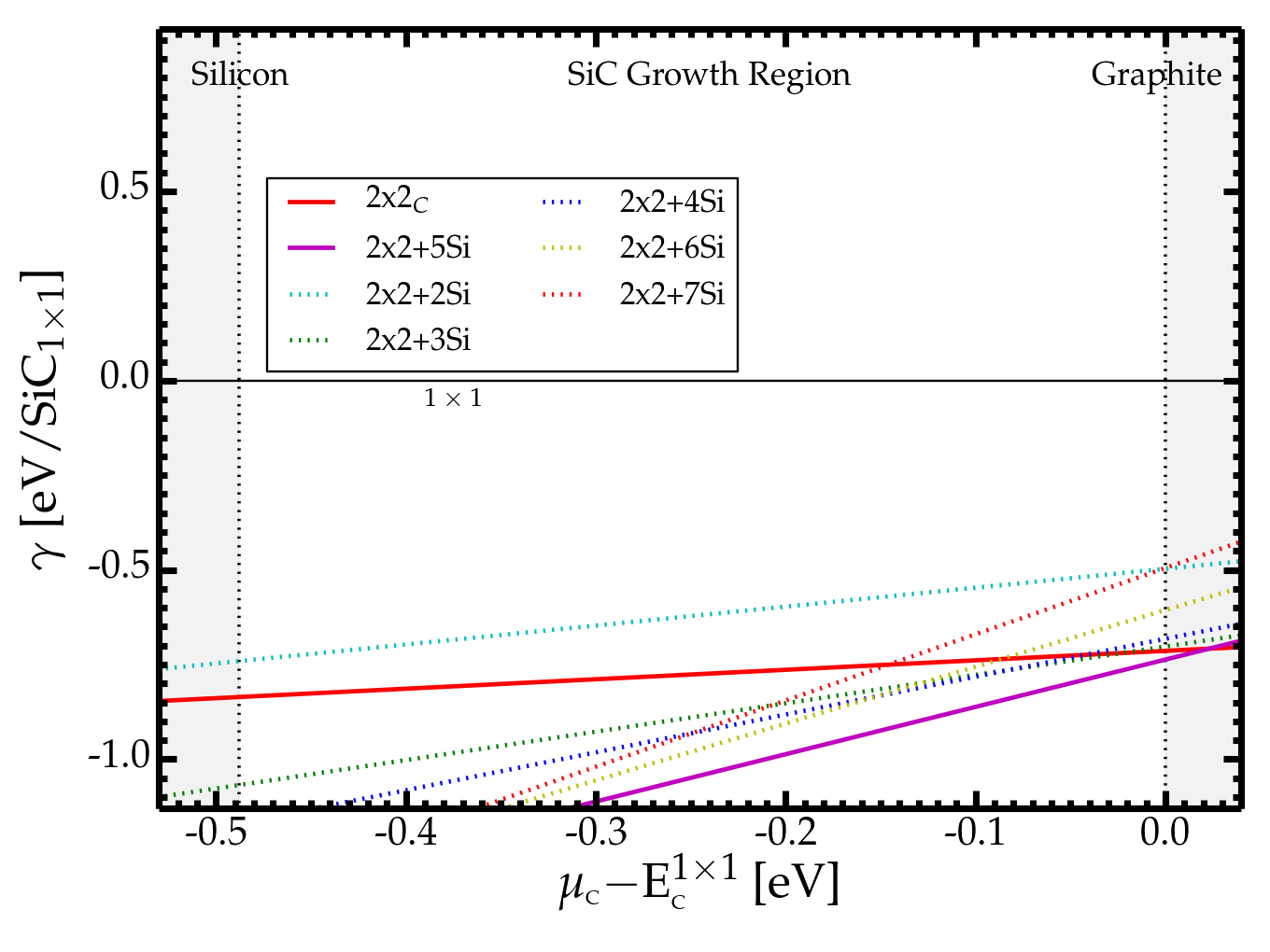}
 \caption{Results for the (2$\times$2) 3C-SiC($\bar{\text{1}}\bar{\text{1}}\bar{\text{1}}$) surface reconstruction composed of silicon adatoms only.
 \label{fig:2x2+si_results}}
\end{figure}

\begin{figure}[ht]
 \includegraphics[width=0.45\textwidth]{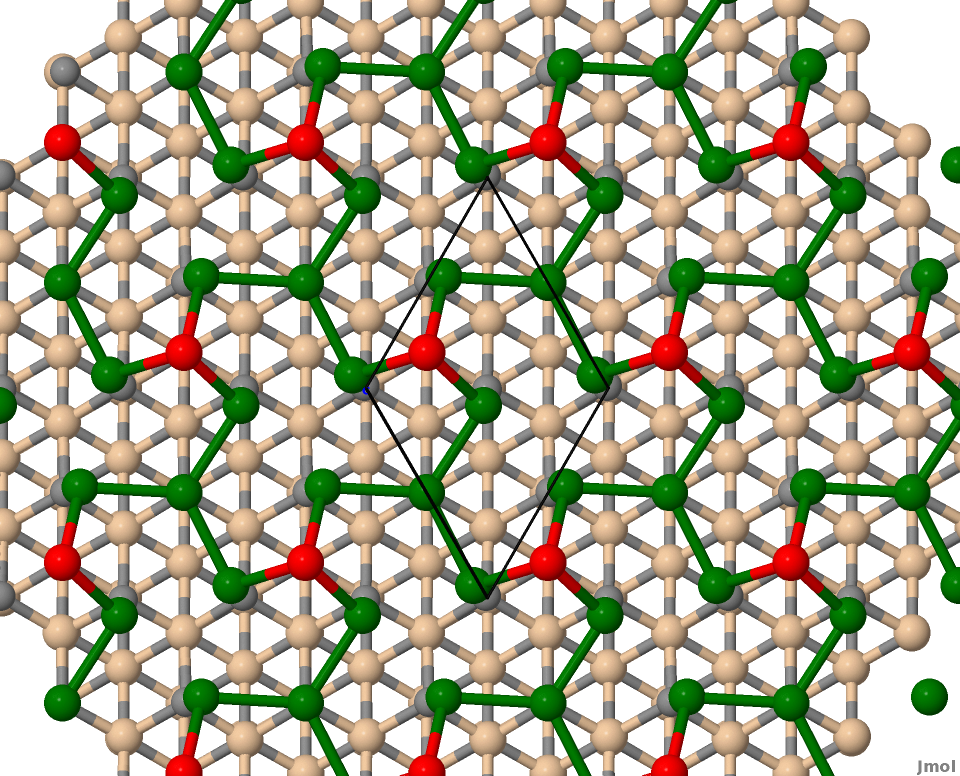}
 \caption{Top view of the atomic structure of the (2$\times$2)+5 Si surface structure - Color: Si monolayer in green, top Si adatoms in red. The unit cell is shown by a black outline.
 \label{fig:geo_2x2+5}}
\end{figure}

Results from the searches of a Si-only reconstruction are shown in Fig.~\ref{fig:2x2+si_results}. The experimentally well known (2$\times$2)$_\textrm{C}$ surface reconstruction is reproduced with our approach and shown in Fig.~\ref{fig:2x2+si_results}. We found a previously experimentally unknown reconstruction that is energetically more favorable than the (2$\times$2)$_\textrm{C}$, labelled (2$\times$2)+5Si. It consists of a silicon monolayer (four atoms) and one silicon adatom on top. A top view of the atomic structure is shown in Fig.~\ref{fig:geo_2x2+5}.

\subsubsection{3C-SiC($\bar{\text{1}}\bar{\text{1}}\bar{\text{1}}$)-(2$\times$2) Si+C Mixed Adatom Models}

\begin{table}[ht]
 \begin{tabular}{c|c|c||c|c|c}
  \hline
  \hline
  Element & adatoms & \# & Element & adatoms & \#\\
  \hline
   C,Si & 1,1 & 1,629 & C,Si & 1,2 & 1,581\\
   C,Si & 1,3 & 1,557 & C,Si & 1,4 & 1,020\\
   C,Si & 2,1 & 1,521 & C,Si & 2,2 & 1,505\\
   C,Si & 2,3 & 1,255 & C,Si & 3,1 & 1,456\\
   C,Si & 3,2 & 1,349 & C,Si & 4,1 & 1,557\\
  \hline
  \hline
 \end{tabular}
 \caption{Total number of relaxed structures for the (2$\times$2) 3C-SiC($\bar{\text{1}}\bar{\text{1}}\bar{\text{1}}$) surface reconstruction with given numbers of silicon and carbon adatoms, where both Si and C are present in the reconstruction (``mixed Si+C''). For the C,Si 2,2 mix the lowest structure to be expected is another BL of silicon carbide and is identified by our approach.\label{tab:2x2_mix_geo_run}}
\end{table}

\begin{figure}[ht]
 \includegraphics[width=0.45\textwidth]{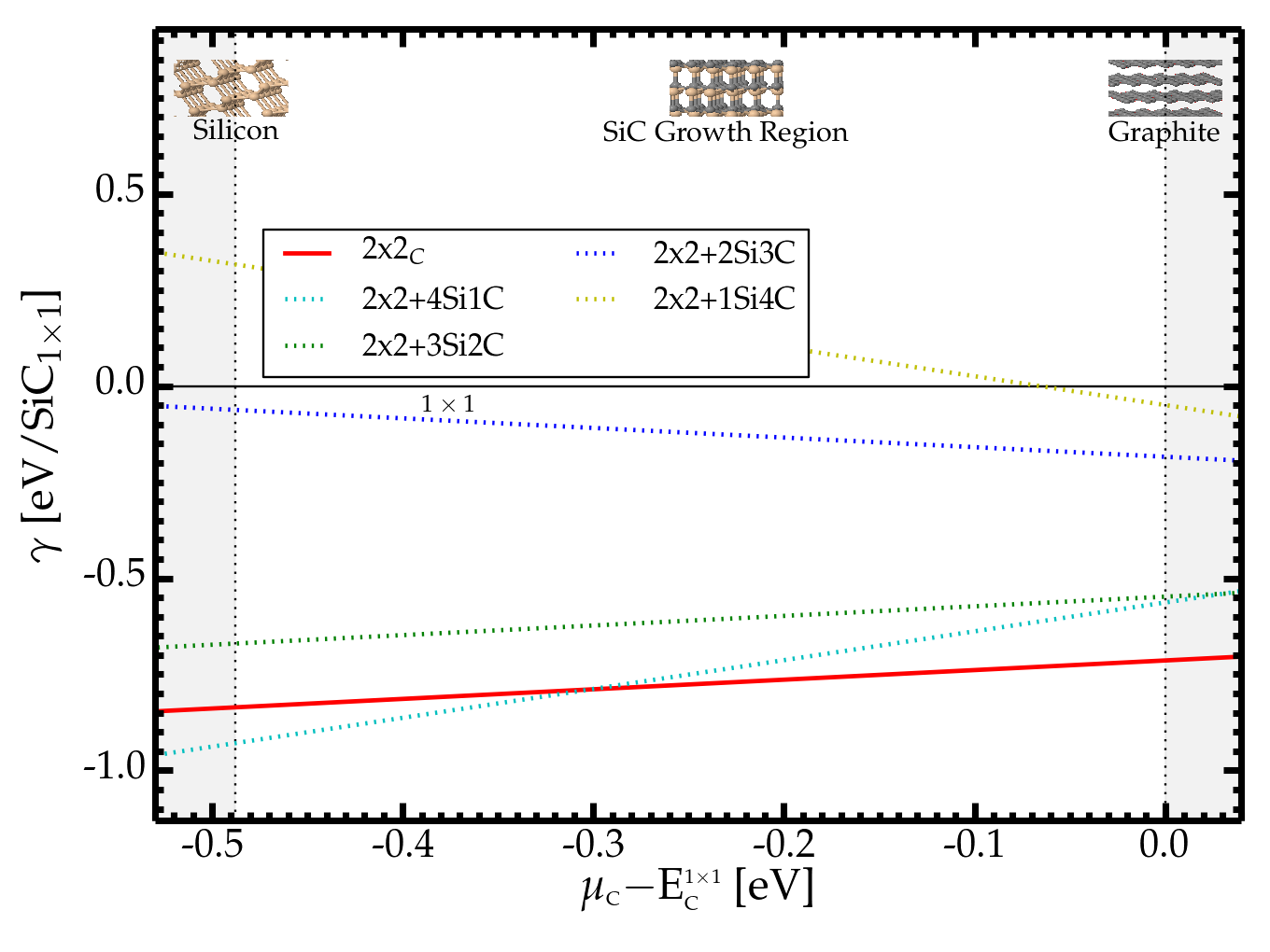}
 \caption{Results for the (2$\times$2) 3C-SiC($\bar{\text{1}}\bar{\text{1}}\bar{\text{1}}$) surface reconstruction composed of a mix of silicon and carbon adatoms.
 \label{fig:2x2_mixes}}
\end{figure}

There were no energetically favorable surface structures from our random searches when adding a mixture silicon and carbon onto the slab. The results are shown in the surface phase diagram for adding silicon + carbon adatoms in Fig.~\ref{fig:2x2_mixes}. The surface energies shown in Fig.~\ref{fig:2x2_mixes} were calculated with Eq.~\ref{eq:surf_energy} using tight settings with PBE+TS and a six-bilayer SiC slab. The surface phase diagram reveals a previously unknown (2$\times$2)+4Si+C, however, the structure is less stable than the (2$\times$2)+5Si structure shown in Fig.~\ref{fig:2x2+si_results}. Our results for the (2$\times$2) surface suggest that the bond between silicon and carbon at the interface between the substrate and the adatoms is energetically favored over the bonding between carbon atoms.

\subsection{The 3C-SiC($\bar{\text{1}}\bar{\text{1}}\bar{\text{1}}$)-(3$\times$3) Surface Reconstruction}

Tables~\ref{tab:3x3_geo_run} and \ref{tab:3x3_geo_run_mix} show a summary of the total number of geometries that were generated and relaxed and the number and type of adatoms that were added to the surface. The maximum number of adatoms per ``layer'' (cf. Fig.~\ref{fig:substrate_with_layers}) on the (3$\times$3) surface is nine, which corresponds to a monolayer covering the surface.

\subsubsection{3C-SiC($\bar{\text{1}}\bar{\text{1}}\bar{\text{1}}$)-(3$\times$3) Si-only and C-only}

\begin{table}[ht]
 \begin{tabular}{c|c|c||c|c|c}
  \hline
  \hline
  Element & adatoms & \# & Element & adatoms & \#\\
  \hline
   C & 1 & 254 & Si & 1 & 1,122\\
   C & 2 & 206 & Si & 2 & 896\\
   C & 3 & 188 & Si & 3 & 847\\
   C & 4 & 169 & Si & 4 & 798\\
   C & 5 & 163 & Si & 5 & 1,689\\
   C & 6 & 145 & Si & 6 & 788\\
   C & 7 & 136 & Si & 7 & 787\\
   C & 8 & 135 & Si & 8 & 728\\
   C & 9 & 128 & Si & 9 & 650\\
   C & 10 & 123 & Si & 10 & 1,370\\
   C & 11 & 108 & Si & 11 & 2,206\\
   C & 12 & 92 & Si & 12 & 1,121\\
   C & 13 & 94 & Si & 13 & 467\\
   C & 14 & 89 & Si & 14 & 449\\
   C & 15 & 67 & Si & 15 & 414\\
  \hline
  \hline
 \end{tabular}
 \caption{Total number of relaxed structures for the (3$\times$3) 3C-SiC($\bar{\text{1}}\bar{\text{1}}\bar{\text{1}}$) surface reconstruction with given numbers of silicon adatoms or C adatoms. The 5, 10, 11 and 12 silicon adatom structures were searched more intensively due to the occurrence of low total energies compared to other structures.
  \label{tab:3x3_geo_run}}
\end{table}

\begin{figure}[h]
 \includegraphics[width=0.45\textwidth]{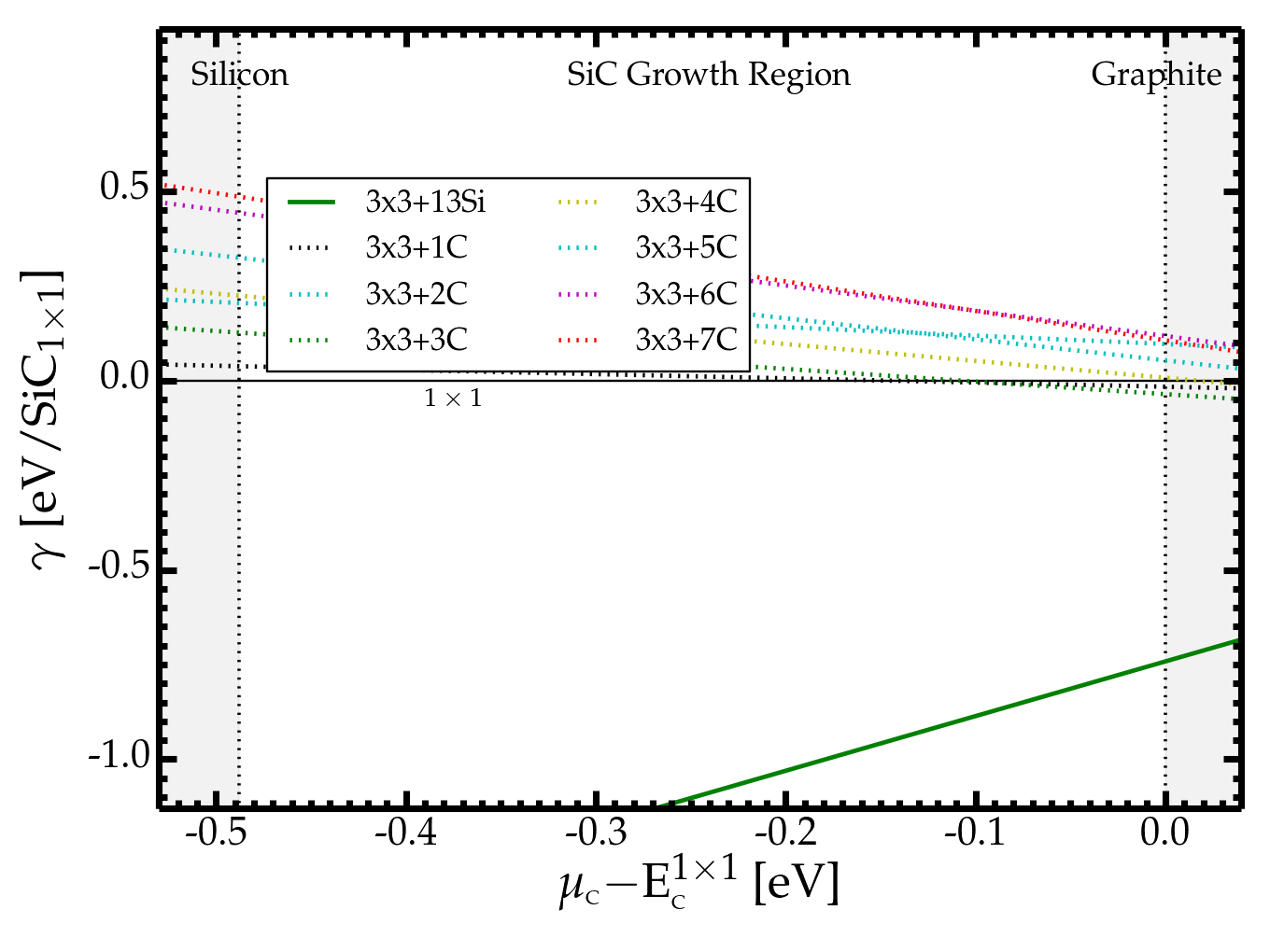}
 \caption{Results from the (3$\times$3) 3C-SiC($\bar{\text{1}}\bar{\text{1}}\bar{\text{1}}$) surface reconstruction composed of carbon adatoms only. For comparison the (3$\times$3)+13~Si (the Si-twist \cite{nemec2015wgg} model) surface structure is included.
 \label{fig:3x3+C}}
\end{figure}

There were no energetically favorable results from our random searches when adding only carbon atoms onto the slab. All C-only adatom surface reconstruction were too high in energy to compete with the previously suggested (3$\times$3)~Si-twist\cite{nemec2015wgg} surface phase. Their energies are shown in the surface phase diagram for adding carbon adatoms only in Fig.~\ref{fig:3x3+C}. The surface energies shown in Fig.~\ref{fig:3x3+C} were calculated using Eq.~\ref{eq:surf_energy}, tight settings,  PBE+TS and a six-bilayer SiC slab underneath.

\begin{figure}[h]
 \includegraphics[width=0.45\textwidth]{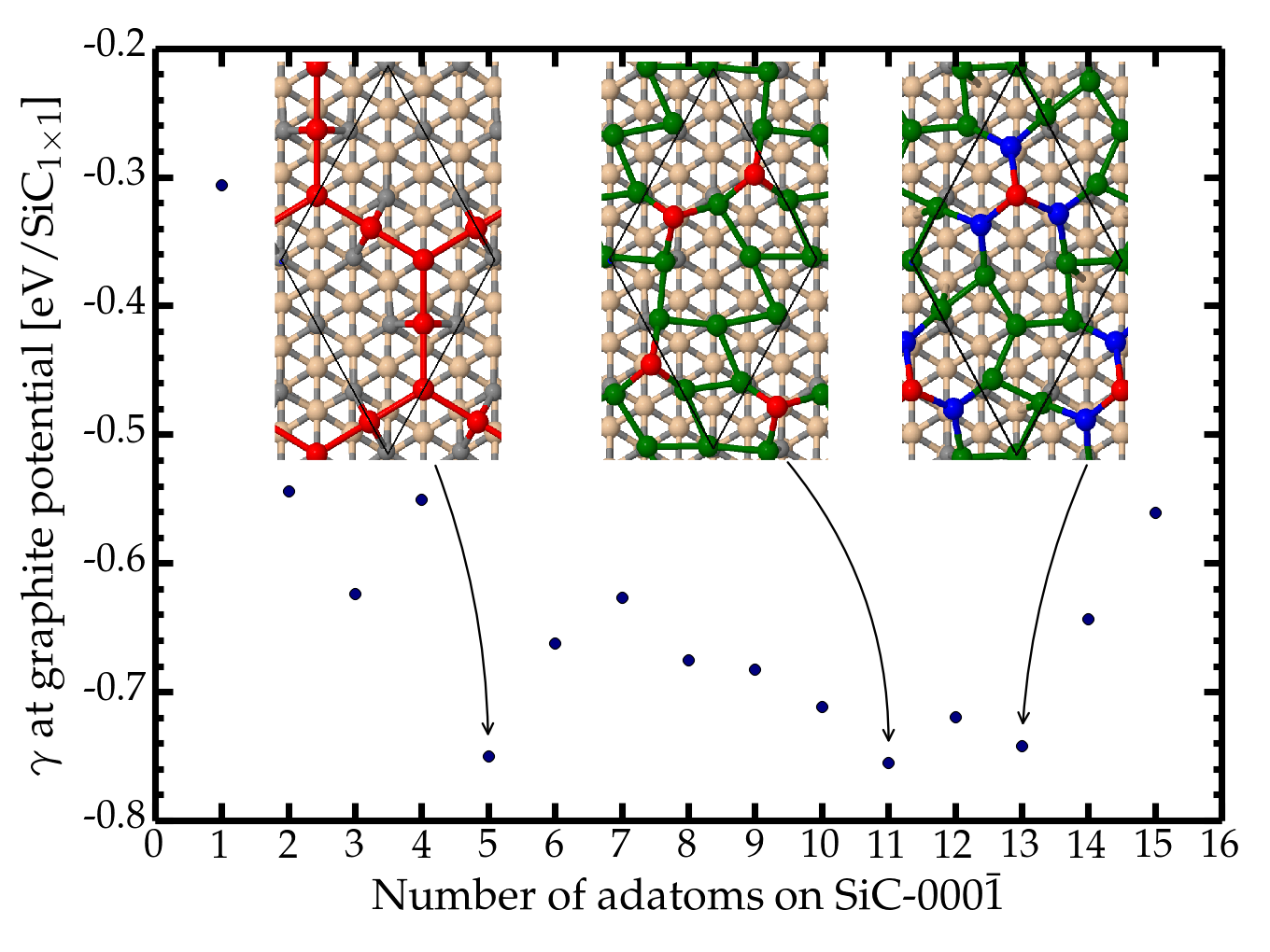}
 \caption{Energetic evolution when successively adding more silicon atoms to the (3$\times$3) 3C-SiC($\bar{\text{1}}\bar{\text{1}}\bar{\text{1}}$) surface. The annotated configurations with lowest total energies are investigated further. Color: Green is a complete silicon monolayer, blue are atoms in between the monolayer and the top adatom, red are top adatoms. Tight settings with PBE+TS.
 \label{fig:3x3_1-15_energies}}
\end{figure}

For the (3$\times$3) 3C-SiC($\bar{\text{1}}\bar{\text{1}}\bar{\text{1}}$) surface reconstruction with silicon adatoms only, we added from 1 to 15 silicon atoms to the surface, performed random structure searches and recomputed the lowest energy candidates of each single class with tight settings PBE+TS and six-bilayer SiC slab. The resulting energies are plotted versus the number of adatoms in Fig.~\ref{fig:3x3_1-15_energies} (identical with main text Fig.~1~(II)) for $\mu_C$ at the graphite line. From Figure~\ref{fig:3x3_1-15_energies}, we derive the numbers of adatoms that are energetically most desirable, namely the (3$\times$3)+5~Si, (3$\times$3)+11~Si and the (3$\times$3)+13~Si model (the Si-twist \cite{nemec2015wgg} model).

\begin{figure}[h]
 \includegraphics[width=0.45\textwidth]{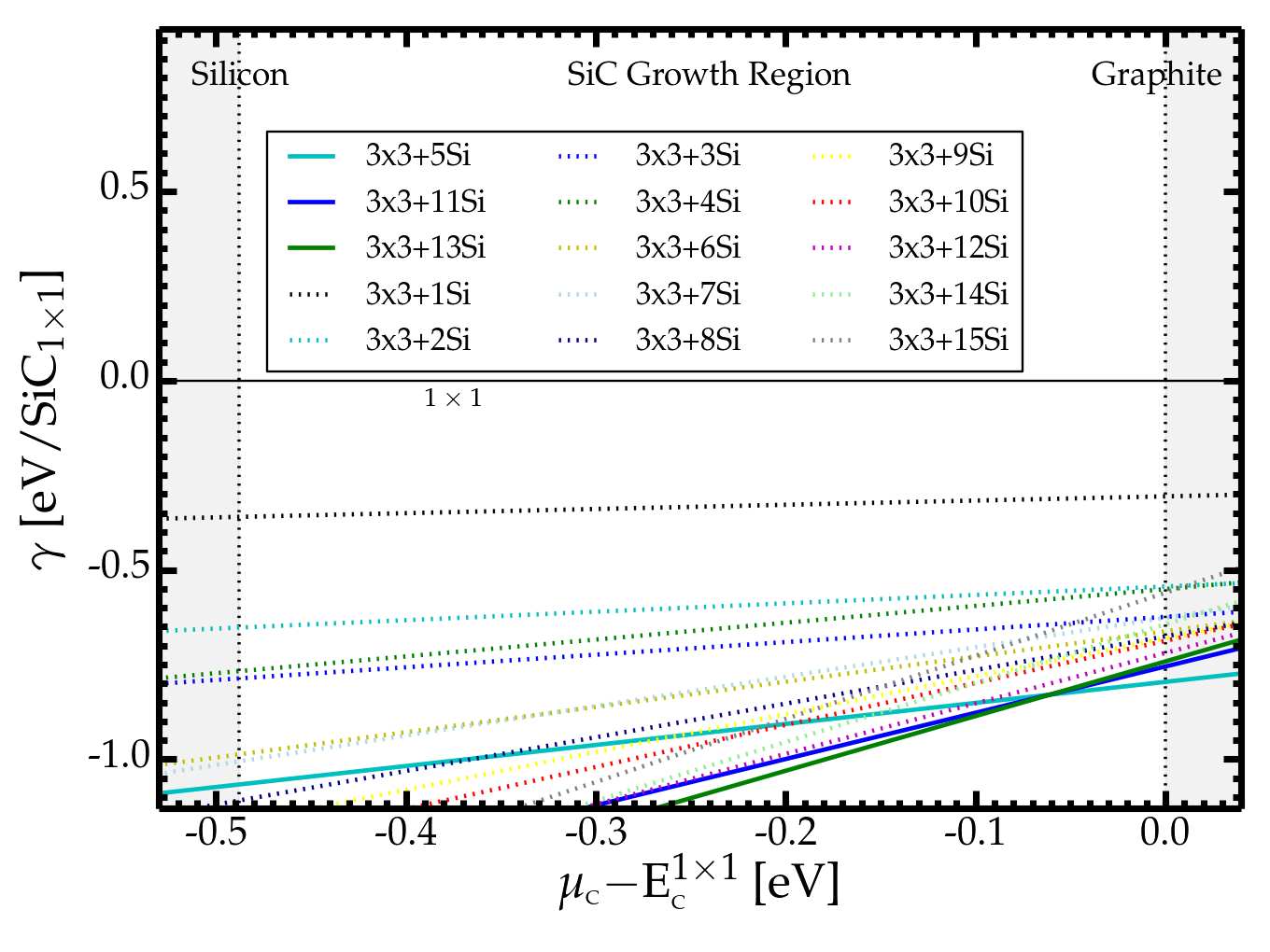}
 \caption{Results from the (3$\times$3) 3C-SiC($\bar{\text{1}}\bar{\text{1}}\bar{\text{1}}$) surface reconstruction composed of silicon adatoms only. The two lowest surface reconstructions are the (3$\times$3)+13Si model (the Si-twist \cite{nemec2015wgg} model) in the Si rich regime and the (3$\times$3)+5Si model in the carbon rich limit. The (3$\times$3)+11Si model is in close competition near the chemical potential where the (3$\times$3) Si-twist model and the (3$\times$3)+5~Si model cross.
 \label{fig:3x3+si}}
\end{figure}

The full surface phase diagram of the low energy structures from Fig.~\ref{fig:3x3_1-15_energies} is shown in Fig.~\ref{fig:3x3+si} using tight settings, PBE+TS and six-bilayer SiC slabs. In the Si-rich limit of the chemical potential, the (3$\times$3)+13~Si (the Si-twist \cite{nemec2015wgg}) is the most stable structure. The (3$\times$3)+5Si model is the most stable surface structure in the carbon rich limit. The (3$\times$3)+11Si model is in close competition near the chemical potential where the energies of the (3$\times$3) Si-twist model and of the  (3$\times$3)+5~Si model cross.

\subsubsection{3C-SiC($\bar{\text{1}}\bar{\text{1}}\bar{\text{1}}$)-(3$\times$3) Si+C Mixed Models}

\begin{table}[ht]
 \begin{tabular}{c|c|c||c|c|c||c|c|c}
  \hline
  \hline
  Element & adat. & \# & Element & adat. & \# & Element & adat. & \#\\
  \hline
   C,Si & 1,1 & 224 & C,Si & 1,2 & 203 & C,Si & 1,3 & 188\\
   C,Si & 1,4 & 185 & C,Si & 1,5 & 182 & C,Si & 1,6 & 185\\
   C,Si & 1,7 & 175 & C,Si & 1,8 & 159 & C,Si & 2,1 & 201\\
   C,Si & 2,2 & 182 & C,Si & 2,3 & 166 & C,Si & 2,4 & 164\\
   C,Si & 2,5 & 166 & C,Si & 2,6 & 154 & C,Si & 2,7 & 165\\
   C,Si & 3,1 & 174 & C,Si & 3,2 & 165 & C,Si & 3,3 & 103\\
   C,Si & 3,4 & 104 & C,Si & 3,5 & 90 & C,Si & 3,6 & 142\\
   C,Si & 4,1 & 160 & C,Si & 4,2 & 153 & C,Si & 4,3 & 103\\
   C,Si & 4,4 & 87 & C,Si & 4,5 & 90 & C,Si & 5,1 & 147\\
   C,Si & 5,2 & 139 & C,Si & 5,3 & 82 & C,Si & 5,4 & 86\\
  \hline
  \hline
 \end{tabular}
  \caption{Total number of relaxed structures for the  3C-SiC($\bar{\text{1}}\bar{\text{1}}\bar{\text{1}}$)-(3$\times$3) surface reconstruction with given numbers of silicon and carbon adatoms that are simultaneously present (``mixed'') at the surface.
  \label{tab:3x3_geo_run_mix}}
\end{table}

\begin{figure}[ht]
 \includegraphics[width=0.45\textwidth]{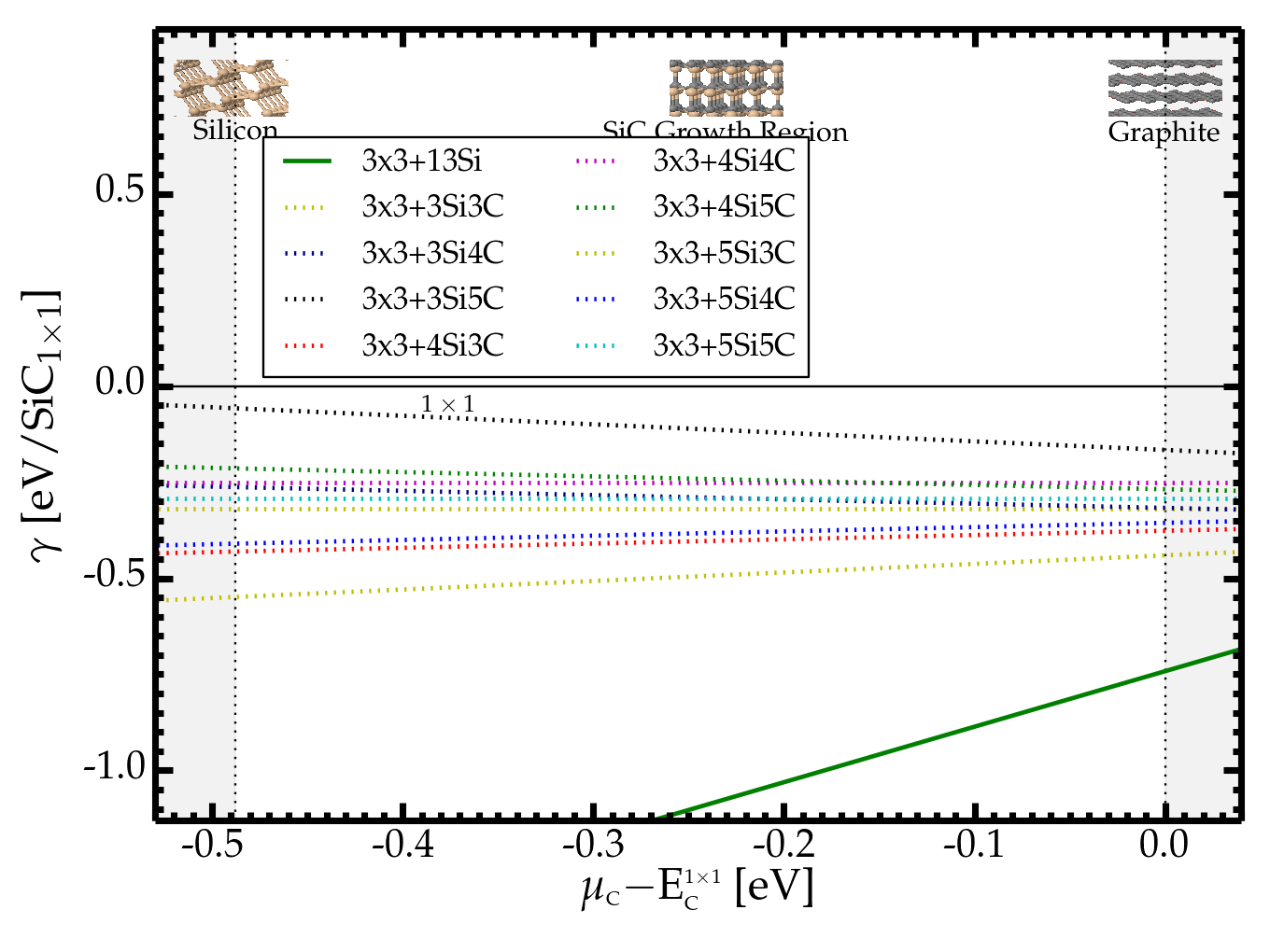}
 \caption{Results for the  3C-SiC($\bar{\text{1}}\bar{\text{1}}\bar{\text{1}}$)-(3$\times$3) surface reconstruction composed of a mix of silicon and carbon adatoms, also showing the previously published (3$\times$3)~Si-twist structure by Nemec \textit{et al.}~\cite{nemec2015wgg} for reference.
 \label{fig:3x3+mix}}
\end{figure}

The were no energetically favorable results from our random searches when adding a mixture of carbon and silicon adatoms onto the slab. The surface phase diagrams shown in Fig.~\ref{fig:3x3+mix} were calculated on the level of tight settings, PBE+TS and a six-bilayer SiC slab. None of the structures included in Fig.~\ref{fig:3x3+mix} were stable over the previously published (3$\times$3)~Si-twist structure by Nemec \textit{et al.}~\cite{nemec2015wgg}.

\section{Simulated Scanning Tunneling Microscopy (STM) Images}
\label{Sec:STM}

Based on the optimized geometries for the low energy structures obtained by the AIRSS approach we simulated scanning tunneling microscope (STM) images to compare against the experimental constant current STM images from Hiebel \textit{et al.}~\cite{hiebel2012sas}
Our approach uses the SIESTA\cite{SIESTA_2002} program and the STM simulation tool written by Pablo Ordej\'on and Nicolas Lorente\cite{siesta_stm_paper}. The simulation of the tip is crucial to obtain comparable STM images and was performed using the STM tip simulation tool described in Ref.\cite{siesta_stm_workers}. The default Si tip was used. We visualized the images using the WSxM\cite{wsxm_program} program.

\begin{figure}[hb]
    \begin{minipage}{0.22\textwidth}
        \includegraphics[width=0.95\textwidth]{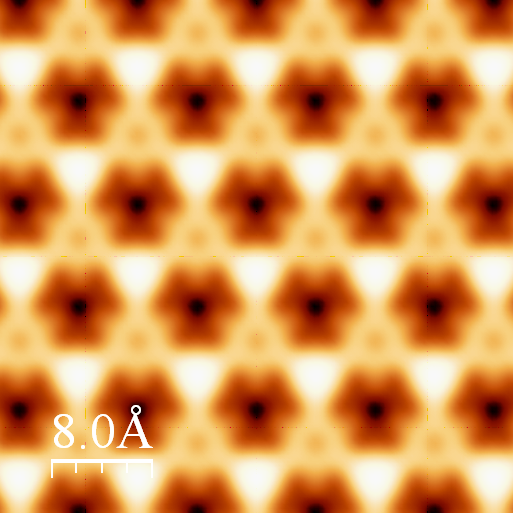}
    \end{minipage}
    \begin{minipage}{0.22\textwidth}
        \includegraphics[width=0.95\textwidth]{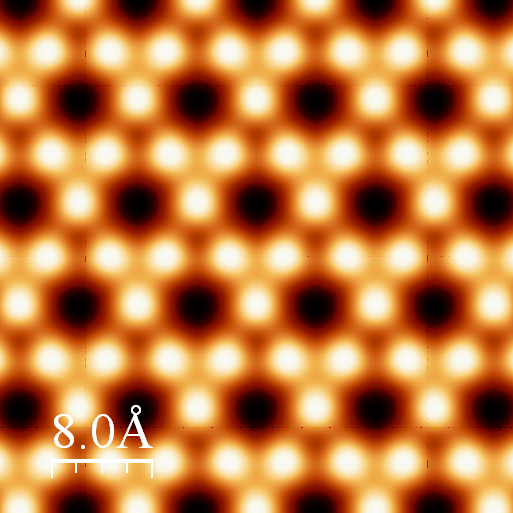}
    \end{minipage}
 \caption{Simulated STM images of the (3$\times$3)+5~Si structure using SIESTA. Shown are the results for a -2.5~V (left) and +2.5~V (right) simulated scanning voltage.\label{fig:STM_posneg_5}}

    \begin{minipage}{0.22\textwidth}
        \includegraphics[width=0.95\textwidth]{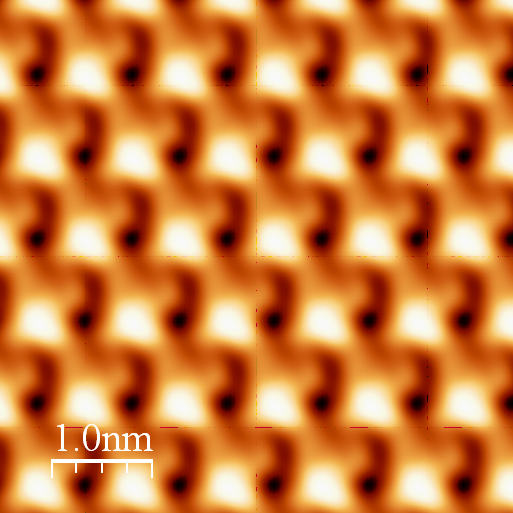}
    \end{minipage}
    \begin{minipage}{0.22\textwidth}
        \includegraphics[width=0.95\textwidth]{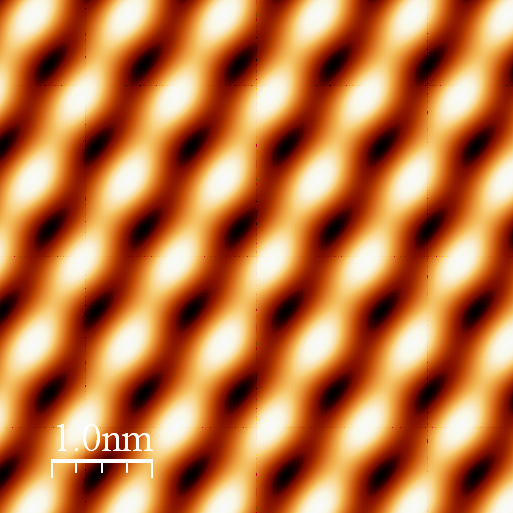}
    \end{minipage}
 \caption{Simulated STM images of the (3$\times$3)+11Si structure using SIESTA. Shown are the results for a -2.5~V (left) and +2.5~V (right) simulated scanning voltage.\label{fig:STM_posneg_11}}

    \begin{minipage}{0.22\textwidth}
        \includegraphics[width=0.95\textwidth]{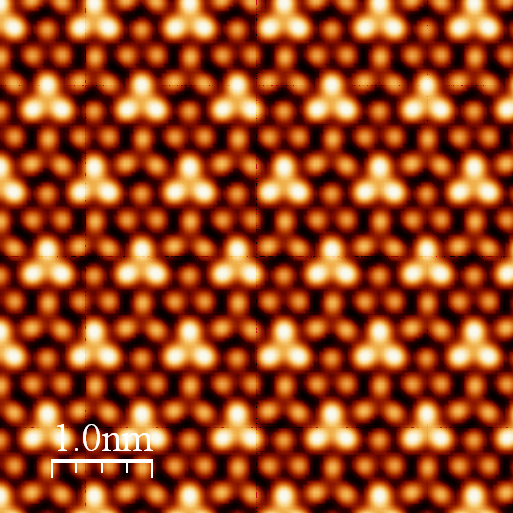}
    \end{minipage}
    \begin{minipage}{0.22\textwidth}
        \includegraphics[width=0.95\textwidth]{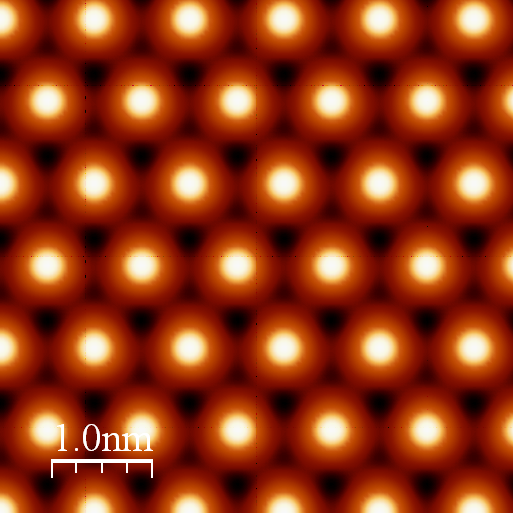}
    \end{minipage}
 \caption{Simulated STM images of the (3$\times$3)~Si-twist structure using SIESTA. Shown are the results for a -2.5~V (left) and a +2.5~V (right) simulated scanning voltage.\label{fig:STM_posneg_13}}
\end{figure}

The computational settings were chosen to match specific experimentally set values of -2.5V and 2.5V, respectively, for which data is reported in Ref.~\cite{hiebel2012sas}. 
As input geometries, we used the three low-energy structure models (3$\times$3))+5~Si (Fig.~\ref{fig:STM_posneg_5}), 3$\times$3)+11~Si (Fig.\ref{fig:STM_posneg_11}), and (3$\times$3)~Si-twist (Fig.\ref{fig:STM_posneg_13}).

The comparison in Figure~2 in the main text between the experimental constant current STM images from Hiebel \textit{et al.}~\cite{hiebel2012sas} and simulated STM images of the (3$\times$3)+5Si shows very good agreement. In Figure~2 in the main text, the three characteristic points  (A, B, C) in the STM images are marked by arrows and labeled according to Hiebel \textit{et al.}~\cite{hiebel2012sas}. In contrast, the low energy structures (3$\times$3)+11~Si and (3$\times$3)~Si-twist reconstruction (Figs.~\ref{fig:STM_posneg_11} and \ref{fig:STM_posneg_13}) fail to describe the characteristic difference in intensity between occupied and empty states that is observed in experimental STM images. 

\section{The SiC/Graphene Interface Structures}

In our previous work, Refs.˜\cite{nemec2013tec} and \cite{nemec2015wgg}, we discuss in detail how to computationally describe SiC/graphene interface structures. To model the interface, we limited our study to a 30$^\circ$ rotation between the substrate and the graphene film, as is observed in experiment on the Si-face of SiC. We here include a short description of the interface calculations. We constructed the interface structures using a ($6 \sqrt{3} \times 6 \sqrt{3}$)-R30$^\circ$ SiC supercell covered by a $\left( 13\times 13 \right)$ graphene cell rotated by 30$^{\circ}$ with respect to the substrate.

The 6$\sqrt{3}$-(2$\times$2)$_{\text{C}}$ interface (labeled (\textit{i}) in Fig.~1(III) in the main text) covers 27 unit cells of the (2$\times$2)$_\text{C}$ reconstruction. Our model of the graphene covered (3$\times$3) phase consist of the same (13$\times$13) graphene supercell, covering 12 units of the (3$\times$3)+5Si (labeled (\textit{k}) in Fig.~1(III) in the main text). As for the surface structures above, the interface structures are calculated using a slab of six SiC bilayers. The bottom silicon atoms are hydrogen terminated. The top three SiC bilayers and all adatoms or planes above are fully relaxed (residual energy gradients: $8 \cdot 10^{-3}$~eV/{\AA} or below).

\newpage

\bibliography{bibliography}